\documentclass[nofootinbib,secnumarabic]{revtex4}

\usepackage{graphicx}
\usepackage{amsmath,amssymb,mathrsfs}
\usepackage{fancyhdr}
\usepackage{psfrag}
\usepackage{array,bbm}
\usepackage{url}
\usepackage{rotating}
\usepackage{multirow}
\usepackage{dsfont}
\usepackage{cancel}
\usepackage{color}
\usepackage{hyperref}
\usepackage{natbib}
\usepackage[utf8]{inputenc}

\usepackage{epsfig}

\numberwithin{equation}{section}
\makeatletter       

\renewcommand{\p@subsection}{}

\makeatother

\newenvironment{Eqnarray}%
     {\arraycolsep 0.14em\begin{eqnarray}}{\end{eqnarray}}
\newcommand{\ba}{\begin{Eqnarray}}
\newcommand{\ea}{\end{Eqnarray}}
\newcommand{\be}{\begin{equation}}
\newcommand{\ee}{\end{equation}}

\begin{document}

\title{The vacuum structure of the Higgs complex singlet-doublet model}

\author{P.M.~Ferreira}
    \email[E-mail: ]{pmmferreira@fc.ul.pt}
\affiliation{Instituto Superior de Engenharia de Lisboa, Portugal}
\affiliation{Centro de F\'{\i}sica Te\'{o}rica e Computacional,
    Universidade de Lisboa, Portugal}

\date{\today}

\begin{abstract}
The complex singlet-doublet model is a popular theory to account for dark matter and
electroweak baryogenesis, wherein the Standard Model particle content is supplemented
by a complex scalar gauge singlet, with certain discrete symmetries imposed. The scalar potential which results thereof can have seven different types of minima at tree-level, which may coexist
for specific choices of parameters. There is therefore the possibility that a given minimum is not
global but rather a local one, and may tunnel to a deeper extremum, thus causing vacuum instability.
This rich vacuum structure is explained and discussed in detail.
\end{abstract}

\maketitle

\section{Introduction}
\label{sec:int}

The recent discovery of the Higgs boson at the LHC \cite{Aad:2012tfa,Chatrchyan:2012xdj}
and subsequent detailed studies of its properties
\cite{Khachatryan:2016vau} have shown that the Standard Model
(SM) is a remarkably successful description of particle physics, even in the scalar sector.
The 125 GeV Higgs boson has properties which, up until now, show no significant deviations
from its SM-expected behaviour. Theoretical extensions of the SM, including new physics --
in particular new particles --
must take that into account, even if there are numerous questions left unexplained by the SM,
such as the matter-antimatter asymmetry, the existence of dark matter or the observed fermion
mass hierarchy. With LHC's Run II about to reveal data on hitherto unexplored energy regions,
generalizations of the SM will finally have a chance to be experimentally tested. One of the
simplest extensions of the SM is the two-Higgs doublet model (2HDM) (\cite{Lee:1973iz}; for a recent
review, \cite{Branco:2011iw}), which has  richer scalar spectrum, potential additional sources
of CP violation via spontaneous symmetry breaking and can also include dark matter candidates.
The 2HDM corresponds to doubling the number of scalar $SU(2)_W$  doublets of the SM, and the
gauge symmetries imposed upon the model simplify considerably the form of the scalar potential.
Still, the potential is a function of eight real scalar component fields (five after electroweak
symmetry breaking), and includes a scalar spectrum composed of five particles (in the most
common versions of the 2HDM, two CP-even scalars, a pseudoscalar and a charged state).

Arguably, the simplest SM extension is that which includes a single extra scalar particle.
That can be achieved by adding a real gauge singlet ({\em i.e.}, a real scalar field with no
$SU(2)_W\times U(1)_Y$ quantum numbers) \cite{McDonald:1993ex,Burgess:2000yq,O'Connell:2006wi,
BahatTreidel:2006kx,Barger:2007im,He:2007tt,Davoudiasl:2004be,Basso:2013nza,Fischer:2016rsh,
Pruna:2013bma,Lopez-Val:2014jva,Robens:2015gla}.
The singlet does not have gauge or fermionic interactions,
and interacts solely with the SM Higgs boson through terms in the scalar potential.
This extra singlet is either a good dark matter candidate or provides a good
mechanism for a strong first order transition explaining baryogenesis, but cannot
accomplish both tasks. The complex singlet-doublet model (cxSM) adds a complex
scalar singlet to the SM particle content, thus the theory has two extra scalars, and
helps explain both the dark matter relic abundance and baryogenesis electroweak phase
transition \cite{Barger:2008jx}. This model has been the subject of many studies
\cite{Dawson:2009yx,Kadastik:2009cu,Lerner:2009xg,Barger:2010yn,Englert:2011yb,
Lebedev:2012zw,Coimbra:2013qq,Costa:2014qga,Costa:2015llh} and has a rich phenomenology.

The stability of the vacuum of the singlet-doublet model has been the subject of some studies
\cite{Gonderinger:2009jp,Lebedev:2012zw,Gonderinger:2012rd,Costa:2014qga}, focussed on the
possible development of instabilities in the scalar potential once quantum corrections
are taken into account -- namely, the one-loop evolution of the potential's quartic couplings
might violate bounded-from-below conditions on those same couplings, much in the same
manner as what may be happening with the SM \cite{Isidori:2001bm,Degrassi:2012ry}.
However, there is another source of potential vacuum instability, deriving from the
fact that, with the addition of extra scalar content, the model's vacuum structure becomes
more complex than the SM's: whereas in the SM, at tree-level, there is only the possibility
of a single
minimum, with extra scalars that is no longer the case, since the additional scalars may,
themselves, acquire a vacuum expectation value (vev). There is thus the possibility
of the potential having more than one minimum for specific combinations of its parameters,
and hence a given minimum is not guaranteed to be the global one. As such, it is
possible that a local minimum tunnels, given enough time, to a deeper minimum. This
situation is to be avoided at all costs -- meaning, the regions of parameter space for
which it may happen should not be considered -- since the consequences of a universal
transition from one minimum to a deeper one would be catastrophic.

In this paper the vacuum structure of the complex singlet-doublet model is analysed in detail,
and it is shown that the model may have a total of {\em seven} different types of stationary points.
A thorough study of these possible vacua will show that there can indeed be simultaneous
minima in the potential already at tree-level, and formulae for the scalar masses, minimization
equations and comparison of the values of the potential at different extrema will be
presented. In section \ref{sec:mod} the model is reviewed, with the potential specified
and its symmetries discussed. The list of possible vacua, and which symmetries are
spontaneously broken are presented, as well as the bounded-from-below conditions that the
quartic couplings of the scalar potential must obey. In section \ref{sec:min} the
minimization conditions, as well as the scalar mass matrices, at each of the seven possible
extrema are presented. In section \ref{sec:sta} the stability, or lack thereof, of each of the
extrema is discussed, and analytical formulae for the difference in depths of the potential
at each extremum are deduced and presented. It will be shown that only one of the seven
possible minima is guaranteed to, if it exists, be the global one. The remainder, as will be
shown, may be local minima and tunnel to lower minima of different types. A numerical scan
over the model's parameter space will be presented in section \ref{sec:con}, to illustrate that
the vacuum instability does not occur for non-interesting corners of parameter space alone,
although it will also be made clear that it is a rare phenomenon. A general discussion is
then held in section \ref{sec:fin}.

\section{The model: symmetries and possible vacua}
\label{sec:mod}

The particle content of the model is identical to the SM's in the fermionic and gauge
sectors, but includes an extended scalar sector. Other than the normal $SU(2)$ doublet
$\Phi$, with hypercharge $Y = 1$, the model also includes a complex gauge singlet ($Y = 0$),
$\chi$. To reduce the large number of parameters of the scalar
potential, two discrete symmetries will be imposed: the potential must be invariant under
a $Z_2$ transformation on the singlet, $\chi \rightarrow -\chi$, and under a CP symmetry,
$\chi \rightarrow \chi^*$. With these requirements, the most general potential is
written as
\ba
V &=& \mu^2_{\Phi}|\Phi|^2\,+\,\mu^2_{\chi, 1} |\chi|^2 \,+\, \mu^2_{\chi, 2}
\left(\chi^2 + \mbox{h.c.}\right)
\,+\, \frac{1}{2}\lambda_\Phi |\Phi|^4 \,+\, \frac{1}{2}\lambda_{\chi, 1} |\chi|^4 \,+\, \lambda_{\chi, 2} \left(\chi^4 + \mbox{h.c.}\right)
\,+\,
\nonumber \\
  && \lambda_{\chi, 3} |\chi|^2\left(\chi^2 + \mbox{h.c.}\right) \,+\, \lambda_{\Phi \chi,1}|\Phi|^2|\chi|^2 \,+\,
 \lambda_{\Phi \chi,2}|\Phi|^2 \left(\chi^2 + \mbox{h.c.}\right) \,,
 \label{eq:Vcomp}
\ea
where all parameters (a total of 9) in the potential are real.

It is of course entirely doable to work with this potential, written in terms of the complex singlet
$\chi$. The works of refs. \cite{Barger:2008jx,Lebedev:2012zw,Englert:2011yb,Coimbra:2013qq,
Costa:2014qga,Costa:2015llh,Darvishi:2016gvm}, for instance,
deal with similar models and follow that procedure.
However, it is perhaps clearer (and with a simpler formalism) to use instead the real
components of that field, $\chi =
(\chi_1\,+\, \mbox{i} \chi_2)/\sqrt{2}$. The main argument for this choice is as follows: the model
contains a complex singlet, and the potential of eq. \eqref{eq:Vcomp} is invariant under a
CP symmetry; it can therefore, in principle, have minima in which the vacuum expectation
value of $\chi$ contains a complex phase. However, as has been shown in chapter 23.7 of
ref. \cite{Branco:1999fs}, there is no CP violation in the scalar sector of this
model with any of its vacua~\footnote{The ultimate reason for this non-existence of
spontaneous CP violation in this model is the fact that the transformation $\chi \rightarrow \chi$
is, for uncharged singlets, as valid a CP transformation as $\chi \rightarrow \chi^*$. And this
alternate definition of CP is preserved by any vacuum, thus no CP violation occurs in this sector.
I thank Lu\'\i s Lavoura for clarifying this point.}.
In fact, to obtain spontaneous CP violation in this model, one has to change its fermionic sector
via the introduction of vector-like quarks (see \cite{Bento:1990wv} and chapter 24.7 of
\cite{Branco:1999fs}).

Thus the possible occurrence of complex phases is a needless complication, since no
interesting CP phenomena occur in the potential due to it. As such, from this point forward,
this model will be discussed in an equivalent manner: a scalar potential containing
a doublet $\Phi$ and {\em two real singlets}, $\chi_1$ and $\chi_2$. Two discrete
symmetries are imposed upon the model: (i) symmetry $S_a$, in which
$\chi_1 \rightarrow -\chi_1$ and $\chi_2 \rightarrow \chi_2$; (ii), symmetry $S_b$, in which
$\chi_1 \rightarrow \chi_1$ and $\chi_2 \rightarrow -\chi_2$. These symmetries are equivalent to
those mentioned above, expressed in terms of the complex singlet $\chi$.
Another argument for dealing
with the potential in terms of two real singlets instead of a complex one is provided
by ref. ~\cite{Kannike:2012pe}, wherein discovering the  bounded-from-below conditions for this
model required the real singlet component formalism.

The lagrangian of the model is thus
\be
{\cal L}\,=\,\left(\partial_\mu \Phi^\dagger\right)\left(\partial^\mu \Phi\right)\,+\,
\frac{1}{2}\partial_\mu \chi_1\partial^\mu \chi_1\,+\,\frac{1}{2}\partial_\mu \chi_2\partial^\mu \chi_2
\,-\,V\,,
\label{eq:lag}
\ee
and the most general scalar potential invariant under the chosen symmetries can be written as
\ba
V &=& \mu_1^2 |\Phi|^2 \,+\, \frac{1}{2}\mu_2^2 \chi_1^2 \,+\, \frac{1}{2}\mu_3^2 \chi_2^2 \,+\,
\frac{\lambda_1}{2} |\Phi|^4 \,+\, \frac{\lambda_2}{8} \chi_1^4 \,+\, \frac{\lambda_3}{8} \chi_2^4 \,+\,
 \frac{1}{2}\lambda_4 |\Phi|^2 \chi_1^2 \,+\, \frac{1}{2}\lambda_5 |\Phi|^2 \chi_2^2 \,+\,
 \frac{1}{4}\lambda_6 \chi_1^2 \chi_2^2 \, ,
 \label{eq:Vr}
\ea
with all $\mu_i$, $\lambda_j$ real, and a total of 9 independent parameters~\footnote{The factors
multiplying
each coupling were chosen for later convenience, and reflect the fact that $\Phi$ is a complex field, and
$\chi_1$, $\chi_2$ are real ones.}. We define the
real components of the doublet $\Phi$ by
\be
\Phi\,=\,\frac{1}{\sqrt{2}}\,\left(
\begin{array}{c} \varphi_1 + \mbox{i} \,\varphi_2 \\ \varphi_3 + \mbox{i} \,\varphi_4 \end{array}
\right).
\label{eq:Phi}
\ee
To emphasize, eq. \eqref{eq:Vr} is simply a different way of writing the potential of eq. \eqref{eq:Vcomp} -- the physics described by both potentials is exactly the same. The potential of ref. \cite{Englert:2011yb},
for instance, included a global $U(1)$ symmetry instead of the two discrete $Z_2$ symmetries
considered here. That potential corresponded to further restrictions on the parameters of the model,
to wit $\mu_2^2 = \mu_3^2$, $\lambda_2 = \lambda_3 = \lambda_6$ and $\lambda_4 = \lambda_5$.
Many other applications of the model consider soft breaking terms as well \cite{Barger:2008jx,
Coimbra:2013qq,Costa:2014qga,Costa:2015llh}.

There are several constraints that must be imposed
on the quartic couplings of the potential for it to be bounded from below -- thus ensuring the existence of
at least one stable minimum for the potential. In particular, it is easy to see
that the coefficients $\lambda_{1,2,3}$ are necessarily positive~\footnote{Throughout this paper
{\em strong stability} conditions will always be considered, in the sense that all inequalities
are taken as strict. Marginal stability could also have been considered -- and one would obtain
conditions such as $\lambda_1\geq 0$, etc. These would require extra conditions being put upon the
quadratic terms for a very small region of parameter space.}:
\be
\lambda_1\,>\,0\;\;\; , \;\;\;\lambda_2\,>\,0\;\;\; , \;\;\;\lambda_3\,>\,0\,.
\label{eq:bfb1}
\ee
An analysis similar to that made for the 2HDM (see, for instance, section 5.7 of
\cite{Branco:2011iw}) leads to the conclusion that further conditions are
necessary, namely
\be
\lambda_{12}\,=\,\lambda_4\,+\,\sqrt{\lambda_1\lambda_2}\,>\,0\;\;\;,\;\;\,
\lambda_{13}\,=\,\lambda_5\,+\,\sqrt{\lambda_1\lambda_3}\,>\,0\;\;\;,\;\;\,
\lambda_{23}\,=\,\lambda_6\,+\,\sqrt{\lambda_2\lambda_3}\,>\,0\,.
\label{eq:bfb2}
\ee
These conditions are not yet sufficient, though. A set of necessary and sufficient
conditions is given by \eqref{eq:bfb1}, \eqref{eq:bfb2} and
finally~\cite{Hadeler,Chang,Kannike:2012pe}
\be
\sqrt{\lambda_1\lambda_2\lambda_3}\,+\,\lambda_4\sqrt{\lambda_3}
\,+\,\lambda_5\sqrt{\lambda_2}\,+\,\lambda_6\sqrt{\lambda_1}\,+\,
\sqrt{2\lambda_{12}\lambda_{13}\lambda_{23}}\,>\,0.
\label{eq:bfb3}
\ee

The potential of eq. \eqref{eq:Vr} can have different extrema, depending on which fields
acquire vevs. On the onset, due to the gauge freedom of the model, it is always possible,
without loss of generality, to align the vev of the doublet with its real and neutral
component. Thus charge breaking is impossible in this model. Depending on which fields
acquire vevs, though, the physics described by each possible vacuum may be markedly
different. In table \ref{tab:vac} the seven possible extrema are listed, with the symmetries
each one of them breaks.
\begin{table*}[ht!]
\caption{\em Possible extrema in the doublet-singlet model. All vevs are real.}
\begin{tabular}{ccccc}
\hline
\hline
Extremum & \hspace{1cm} & Vevs & \hspace{1cm}  & Symmetries Broken \\
\hline
 & & & & \\
A & & $\langle\Phi\rangle \neq 0\,,\, \langle\chi_1\rangle = 0 \,,\, \langle\chi_2\rangle = 0$ & &
$SU(2)_W\times U(1)_Y$  \\
 & & & &  \\
B & & $\langle\Phi\rangle \neq 0\,,\, \langle\chi_1\rangle \neq 0 \,,\, \langle\chi_2\rangle = 0$ & &
$SU(2)_W\times U(1)_Y$ and $S_a$ \\
 & & & & \\
C & & $\langle\Phi\rangle \neq 0\,,\, \langle\chi_1\rangle = 0 \,,\, \langle\chi_2\rangle \neq 0$ & &
$SU(2)_W\times U(1)_Y$ and $S_b$ \\
 & & & & \\
D & & $\langle\Phi\rangle \neq 0\,,\, \langle\chi_1\rangle \neq 0 \,,\, \langle\chi_2\rangle \neq 0$ & &
$SU(2)_W\times U(1)_Y$, $S_a$ and $S_b$ \\
 & & & & \\
E & & $\langle\Phi\rangle = 0\,,\, \langle\chi_1\rangle \neq 0 \,,\, \langle\chi_2\rangle = 0$ & &
$S_a$ \\
 & & & & \\
F & & $\langle\Phi\rangle = 0\,,\, \langle\chi_1\rangle = 0 \,,\, \langle\chi_2\rangle \neq 0$ & &
$S_b$ \\
 & & & & \\
G & & $\langle\Phi\rangle = 0\,,\, \langle\chi_1\rangle \neq 0 \,,\, \langle\chi_2\rangle \neq 0$ & &
$S_a$ and $S_b$ \\
 & & & & \\
\hline
\hline
\end{tabular}
\label{tab:vac}
\end{table*}

Electroweak symmetry breaking only occurs for extrema of types $A$ to $D$ - only in those situations
can the Higgs mechanism give mass to elementary particles. As such, eventual minima $E$, $F$ and $G$
are completely unphysical and should be avoided at all costs. On the other hand, eventual vacua
of types $A$ to $D$ would have different phenomenologies, due to the broken/unbroken symmetries
$S_a$, $S_b$: such unbroken symmetries mean that some of the scalar fields do not mix with each
other, do not couple at all to fermions or gauge bosons and as such the model has spin-0 dark matter
candidates. In short:
\begin{itemize}
\item A vacuum of type $A$ has a scalar identical to the SM Higgs; the extra scalars,
$\chi_1$ and $\chi_2$, do not mix with the SM-like Higgs and are thus dark matter candidates.
The ``dark sector" only couples to the SM-like scalar via quartic couplings in the potential.
\item In a vacuum of type $B$, the real, neutral component of $\Phi$, $\varphi_3$, mixes
with the real singlet $\chi_1$, thus originating two neutral scalars which couple to fermions
and gauge bosons. The remaining singlet $\chi_2$ would behave as dark matter.
\item A vacuum of type $C$ is very similar to vacuum $B$: now $\varphi_3$ mixes
with $\chi_2$, and it would be $\chi_1$ who would behave as dark matter.
\item In a vacuum of type $D$, the fields $\varphi_3$, $\chi_1$ and $\chi_2$ all mix with
each other, originating three neutral scalars with couplings to fermions and gauge bosons.
No dark matter candidate exists.
\end{itemize}
It could be argued that some of these extrema are related to others -- for instance,
extrema $B$ and $C$ differ only on which singlet, $\chi_1$ or $\chi_2$, acquire a
vev. A basis change of these two
fields~\footnote{For this model, a basis change consists of a generic rotation between
the fields $\chi_1$ and $\chi_2$, since they are, for all intents and purposes, physically
indistinguishable {\em a priori}. The basis changes, of course, cannot mix fields with different
quantum numbers, and as such do not involve the doublet components. In passing, it should be
noted that the arguments made here at this point would also apply to simultaneous
extrema $E$ and $F$.} could be performed to transform
extrema $B$ into a new extrema of type $C$, in the new basis. However, as will be seen
in section \ref{sec:sta}, that is not the case -- on a given, fixed basis of fields,
there is actually the possibility of extrema $B$ and $C$ having physical impact on
the stability of the model. Also, if on a given basis of fields there exist
extrema $B$ and $C$, it will certainly be possible to ``rotate" the $\chi_1$ vev from
extremum $B$ to the field $\chi_2$ via a suitable basis transformation -- thus, in
the new basis, the old extremum $B$ would become an extremum of type $C$. However, bear
in mind that that selfsame basis transformation would change the form of the old
extremum $C$ -- in all likelyhood, the basis-transformed extremum $C$ would have
vevs in both $\chi_1$ or $\chi_2$, and thus be of type $D$.

All of the extrema of table \ref{tab:vac} are therefore possible minima of the potential
of eq. \eqref{eq:Vr}, depending on the values of its nine parameters. That then begs the
question, does this model have only global minima, or can it have local ones? Is it possible
for a physically interesting minimum (say, type $A$) to coexist with a deeper unphysical
global minimum (say, type $F$)? If so, the model's stability would not be guaranteed.
As will be shown in section~\ref{sec:sta}, this situation may well happen, which will
allow to constrain the model's parameter space. First, though, each possible vacuum should
be thoroughly analysed.

\section{Minimization conditions and mass matrices}
\label{sec:min}

In this section the implications for the (real) vevs of the minimization conditions in each
of the possible extrema of table \ref{tab:vac} will be analysed. Also, the scalar mass matrices
will be computed for each of those extrema.

It will be useful to introduce a bilinear notation similar to that used for the study of the
two-Higgs doublet model vacuum structure~\cite{Velhinho:1994np,Ferreira:2004yd,Barroso:2005sm,
Nishi:2006tg,Maniatis:2006fs,Ivanov:2006yq,Barroso:2007rr,Nishi:2007nh,Maniatis:2007vn,
Ivanov:2007de,Maniatis:2007de,Nishi:2007dv,Maniatis:2009vp,Ferreira:2010hy}. The bilinear
formalism simplifies considerably the search for relations between different extrema, and one
defines
\be
x_1 = |\Phi|^2 \;\;\; , \;\;\; x_2 = \frac{1}{2}\chi_1^2 \;\;\; , \;\;\; x_3 = \frac{1}{2}\chi_2^2
\ee
as well as the matrices
\be
X\,=\, \left(\begin{array}{c} x_1 \\ x_2 \\ x_3 \end{array}\right)\;\;\; , \;\;\;
A\,=\, \left(\begin{array}{c} \mu_1^2 \\ \mu_2^2 \\ \mu_3^2 \end{array}\right)\;\;\; , \;\;\;
B\,=\, \left(\begin{array}{ccc} \lambda_1 & \lambda_4 & \lambda_5 \\
\lambda_4 & \lambda_2 & \lambda_6 \\ \lambda_5 & \lambda_6 & \lambda_3 \end{array}
\right)\;\;\; .
\label{eq:def}
\ee
With these definitions, the scalar potential of eq. \eqref{eq:Vr} is written simply as
\be
V\,=\, A^T X \,+\, \frac{1}{2}\,X^T B X\,.
\ee
It is simple to verify that, at a given stationary point for which the fields acquires
vacuum expectation values such that $\langle X \rangle \,=\, (\langle |\Phi| \rangle^2\,,\,\langle \chi_1 \rangle^2/2\,,\,
\langle \chi_2 \rangle^2/2)^T$, the value of the potential at that stationary points, $V_{SP}$,
can be expressed as
\be
V_{SP} \,=\,\frac{1}{2}\,A^T \langle X \rangle \,=\,-\,\frac{1}{2}\,\langle X \rangle^T B \langle X \rangle\,.
\label{eq:Vmin}
\ee

A word on the scalar mass matrices obtained at the several extrema: since the vev of the doublet,
if it is non zero, is aligned with the $\varphi_3$ component, the $\varphi_1$, $\varphi_2$
and $\varphi_4$ components of $\Phi$ do not mix with the others -- rather, they constitute
the model's Goldstone bosons when the electroweak symmetry is broken. For extrema of types $E$ to
$G$, electroweak symmetry breaking does not occur and all four components of the doublet have
the same mass. As such, the masses of neutral scalars of interest in each extremum are
the eigenvalues of a $3\times 3$ matrix, representing eventual mixings between $\varphi_3$, $\chi_1$
and $\chi_2$. With the conventions of eqs.~\eqref{eq:lag} and~\eqref{eq:Phi}, the squared mass matrix
of the neutral scalars will be given by
\be
\left[M^2\right]_{ij}\,=\,\frac{\partial^2 V}{\partial f_i \partial f_j}
\label{eq:mm}
\ee
with $f_{1,2,3} \,=\, \{\varphi_3, \chi_1, \chi_2\}$.

\subsection{Vacuum of type $A$}
\label{sec:minA}

In an extremum of type $A$, according to table \ref{tab:vac}, only the doublet $\Phi$
acquires a vev:
\be
\langle \Phi\rangle_A\,=\,\frac{v_A}{\sqrt{2}}\;\;\; , \;\;\; \langle \chi_1\rangle_A\,=\,0
\;\;\; , \;\;\; \langle \chi_2\rangle_A\,=\,0\, ,
\ee
and the only non-trivial stationarity condition of the potential is
\be
\frac{\partial V}{\partial \varphi_3}\,=\,0 \,\Leftrightarrow \, v_A
\left( \mu_1^2\,+\, \frac{1}{2}\,\lambda_1\,v_A^2\right)\,=\,0.
\label{eq:minA}
\ee
Simple calculations show that the neutral scalar squared mass matrix \eqref{eq:mm} is
given, for this vacuum, by
\be
\left[M^2\right]_A\,=\,\left(\begin{array}{ccc}
\mu_1^2\,+\,\frac{3}{2}\,\lambda_1\,v_A^2 & 0 & 0 \\
0 & \mu_2^2\,+\,\frac{1}{2}\,\lambda_4\,v_A^2 & 0 \\
0 & 0 & \mu_3^2\,+\,\frac{1}{2}\,\lambda_5\,v_A^2  \\
\end{array}\right)
\ee
With the minimization condition \eqref{eq:minA}, the three squared neutral masses at this
stationary point are therefore given by
\be
m^2_{A1}\,=\,\lambda_1\,v_A^2\;\;\; , \;\;\; m^2_{A2}\,=\,\mu_2^2\,+\,\frac{1}{2}\,\lambda_4\,v_A^2
\;\;\; , \;\;\; m^2_{A3}\,=\,\mu_3^2\,+\,\frac{1}{2}\,\lambda_5\,v_A^2\,.
\label{eq:maA}
\ee
Since electroweak symmetry is broken, the fields $\varphi_1$, $\varphi_2$ and $\varphi_4$ are the
Goldstone bosons of the theory. An extremum of type $A$ is therefore a minimum if and only if
eq. \eqref{eq:minA} admits a solution and all squared masses in eq. \eqref{eq:maA} are positive.

For later convenience, one defines the following two vectors:
\be
X_A\,=\,\langle X\rangle_A\,=\,\frac{1}{2}\,\left(\begin{array}{c} v_A^2 \\ 0 \\ 0 \end{array}\right)
\;\;\; , \;\;\;
V^\prime_A \,=\, A \,+\,B\,X_A \,=\, \left(\begin{array}{c} 0 \\ m^2_{A2} \\ m^2_{A3} \end{array}\right)\, .
\label{eq:VlA}
\ee
The components of $V^\prime_A$ are dictated by the minimization condition \eqref{eq:minA}
and the mass definitions of \eqref{eq:maA}.

\subsection{Vacuum of type $B$}
\label{sec:minB}

In an extremum of type $B$, according to table \ref{tab:vac}, the doublet $\Phi$ and the singlet
$\chi_1$ acquire a vev:
\be
\langle \Phi\rangle_B\,=\,\frac{v_B}{\sqrt{2}}\;\;\; , \;\;\; \langle \chi_1\rangle_B\,=\,w_B
\;\;\; , \;\;\; \langle \chi_2\rangle_B\,=\,0\, ,
\ee
and the only non-trivial stationarity conditions of the potential are given by
\ba
\frac{\partial V}{\partial \varphi_3}\,=\,0 &\Leftrightarrow & v_B
\left( \mu_1^2\,+\, \frac{1}{2}\,\lambda_1\,v_B^2\,+\,\frac{1}{2}\,\lambda_4\,w_B^2\right)\,=\,0.
\nonumber \\
\frac{\partial V}{\partial \chi_1}\,=\,0 &\Leftrightarrow & w_B
\left( \mu_2^2\,+\, \frac{1}{2}\,\lambda_4\,v_B^2\,+\,\frac{1}{2}\,\lambda_2\,w_B^2\right)\,=\,0.
\label{eq:minB}
\ea
The neutral scalar squared mass matrix is therefore given by
\be
\left[M^2\right]_B\,=\,\left(\begin{array}{ccc}
\lambda_1\,v_B^2 & \lambda_4 \,v_B \,w_B & 0 \\
\lambda_4 \,v_B \,w_B & \lambda_2\,w_B^2 & 0 \\
0 & 0 & \mu_3^2\,+\,\frac{1}{2}\,\lambda_5\,v_B^2 \,+\,\frac{1}{2}\,\lambda_6\,w_B^2 \\
\end{array}\right)
\ee
The three squared neutral masses at this stationary point are therefore given by
\be
m^2_{B1,2}\,=\,\frac{1}{2}\left[ \lambda_1\,v_B^2\,+\,\lambda_2\,w_B^2\,\pm
\sqrt{\left(\lambda_1\,v_B^2\,-\,\lambda_2\,w_B^2\right)^2\,+\,4\,\lambda_4^2\,v_B^2\,w_B^2}\,\right]
\;\;\; , \;\;\;
m^2_{B3}\,=\,\mu_3^2\,+\,\frac{1}{2}\,\lambda_5\,v_B^2 \,+\,\frac{1}{2}\,\lambda_6\,w_B^2\,.
\label{eq:maB}
\ee
Again, the fields $\varphi_1$, $\varphi_2$ and $\varphi_4$ will be massless and are the
Goldstone bosons of the theory. An extremum of type $B$ is therefore a minimum if and only if
eqs. \eqref{eq:minB} admits a solution and all squared masses in eq. \eqref{eq:maB} are positive.

For later convenience, one defines the following two vectors:
\be
X_B\,=\,\langle X\rangle_B\,=\,\frac{1}{2}\,\left(\begin{array}{c} v_B^2 \\ w_B^2 \\ 0 \end{array}\right)
\;\;\; , \;\;\;
V^\prime_B \,=\, A \,+\,B\,X_B \,=\, \left(\begin{array}{c} 0 \\ 0 \\ m^2_{B3} \end{array}\right)\, .
\label{eq:VlB}
\ee
The components of $V^\prime_B$ are dictated by the minimization conditions \eqref{eq:minB}
and the mass definitions of \eqref{eq:maB}.

\subsection{Vacuum of type $C$}
\label{sec:minC}

In an extremum of type $C$, according to table \ref{tab:vac}, the doublet $\Phi$ and the singlet
$\chi_2$ acquire a vev. We obtain results analogous to those of extremum $B$,
\be
\langle \Phi\rangle_C\,=\,\frac{v_C}{\sqrt{2}}\;\;\; , \;\;\; \langle \chi_1\rangle_C\,=\,0
\;\;\; , \;\;\; \langle \chi_2\rangle_C\,=\,z_C\, ,
\ee
and the only non-trivial stationarity conditions of the potential are given by
\ba
\frac{\partial V}{\partial \varphi_3}\,=\,0 &\Leftrightarrow & v_C
\left( \mu_1^2\,+\, \frac{1}{2}\,\lambda_1\,v_C^2\,+\,\frac{1}{2}\,\lambda_5\,z_C^2\right)\,=\,0.
\nonumber \\
\frac{\partial V}{\partial \chi_2}\,=\,0 &\Leftrightarrow & z_C
\left( \mu_3^2\,+\, \frac{1}{2}\,\lambda_5\,v_C^2\,+\,\frac{1}{2}\,\lambda_3\,z_C^2\right)\,=\,0.
\label{eq:minC}
\ea
The neutral scalar squared mass matrix is therefore given by
\be
\left[M^2\right]_C\,=\,\left(\begin{array}{ccc}
\lambda_1\,v_C^2 & 0 & \lambda_5 \,v_C \,z_C \\
0 & \mu_2^2\,+\,\frac{1}{2}\,\lambda_4\,v_C^2 \,+\,\frac{1}{2}\,\lambda_6\,z_C^2 & 0 \\
\lambda_5 \,v_C \,z_C & 0 & \lambda_3\,z_C^2 \\
\end{array}\right)
\ee
The three squared neutral masses at this stationary point are therefore given by
\be
m^2_{C1,3}\,=\,\frac{1}{2}\left[ \lambda_1\,v_C^2\,+\,\lambda_3\,z_C^2\,\pm
\sqrt{\left(\lambda_1\,v_C^2\,-\,\lambda_3\,z_C^2\right)^2\,+\,4\,\lambda_5^2\,v_C^2\,z_C^2}\,\right]
\;\;\; , \;\;\;
m^2_{C2}\,=\,\mu_2^2\,+\,\frac{1}{2}\,\lambda_4\,v_C^2 \,+\,\frac{1}{2}\,\lambda_6\,z_C^2\,.
\label{eq:maC}
\ee
Again, the fields $\varphi_1$, $\varphi_2$ and $\varphi_4$ will be massless and are the
Goldstone bosons of the theory. An extremum of type $B$ is therefore a minimum if and only if
eqs. \eqref{eq:minC} admits a solution and all squared masses in eq. \eqref{eq:maC} are positive.

For later convenience, one defines the following two vectors:
\be
X_C\,=\,\langle X\rangle_C\,=\,\frac{1}{2}\,\left(\begin{array}{c} v_C^2 \\ 0 \\ z_C^2 \end{array}\right)
\;\;\; , \;\;\;
V^\prime_C \,=\, A \,+\,B\,X_C \,=\, \left(\begin{array}{c} 0 \\ m^2_{C2} \\ 0 \end{array}\right)\, .
\label{eq:VlC}
\ee
The components of $V^\prime_C$ are dictated by the minimization conditions \eqref{eq:minC}
and the mass definitions of \eqref{eq:maC}.

\subsection{Vacuum of type $D$}
\label{sec:minD}

In an extremum of type $D$, according to table \ref{tab:vac}, all three fields acquire a vev,
\be
\langle \Phi\rangle_D\,=\,\frac{v_D}{\sqrt{2}}\;\;\; , \;\;\; \langle \chi_1\rangle_D\,=\,w_D
\;\;\; , \;\;\; \langle \chi_2\rangle_C\,=\,z_D\, .
\ee
The stationarity conditions of the potential are
\ba
\frac{\partial V}{\partial \varphi_3}\,=\,0 &\Leftrightarrow & v_D
\left( \mu_1^2\,+\, \frac{1}{2}\,\lambda_1\,v_D^2\,+\,\frac{1}{2}\,\lambda_4\,w_D^2\,+\,\frac{1}{2}\,\lambda_5\,z_D^2\right)\,=\,0.
\nonumber \\
\frac{\partial V}{\partial \chi_1}\,=\,0 &\Leftrightarrow & w_D
\left( \mu_2^2\,+\, \frac{1}{2}\,\lambda_4\,v_D^2\,+\,\frac{1}{2}\,\lambda_2\,w_D^2\,+\,\frac{1}{2}\,\lambda_6\,z_D^2\right)\,=\,0.
\nonumber \\
\frac{\partial V}{\partial \chi_2}\,=\,0 &\Leftrightarrow & z_D
\left( \mu_3^2\,+\, \frac{1}{2}\,\lambda_5\,v_D^2\,+\,\frac{1}{2}\,\lambda_6\,w_D^2\,+\,\frac{1}{2}\,\lambda_3\,z_D^2\right)\,=\,0.
\label{eq:minD}
\ea
The neutral scalar squared mass matrix is
\be
\left[M^2\right]_D\,=\,\left(\begin{array}{ccc}
\lambda_1\,v_D^2 & \lambda_4 \,v_D \,w_D & \lambda_5 \,v_D \,z_D \\
\lambda_4 \,v_D \,w_D & \lambda_2\,w_D^2 & \lambda_6 \,w_D \,z_D \\
\lambda_5 \,v_D \,z_D & \lambda_6 \,w_D \,z_D & \lambda_3\,z_D^2 \\
\end{array}\right)\;\;\; = \;\;\; \left(\begin{array}{ccc} v_D & 0 & 0 \\
 0 & w_D & 0 \\ 0 & 0 & z_D \end{array}\right)\;B\;
 \left(\begin{array}{ccc} v_D & 0 & 0 \\
 0 & w_D & 0 \\ 0 & 0 & z_D \end{array}\right)\,.
\label{eq:maD}
\ee
The three squared neutral masses at this stationary point ($m^2_{D1}$, $m^2_{D2}$, $m^2_{D3}$)
are the eigenvalues of this simple matrix. When $D$ is a minimum, clearly, eq. \eqref{eq:maD}
implies that the matrix $B$ is positive definite. And again, the fields $\varphi_1$, $\varphi_2$ and
$\varphi_4$ will be massless and are the Goldstone bosons of the theory.

For later convenience, one defines the following two vectors:
\be
X_D\,=\,\langle X\rangle_D\,=\,\frac{1}{2}\,\left(\begin{array}{c} v_D^2 \\ w_D^2 \\ z_D^2 \end{array}\right)
\;\;\; , \;\;\;
V^\prime_D \,=\, A \,+\,B\,X_D \,=\, \left(\begin{array}{c} 0 \\ 0 \\ 0 \end{array}\right)\, .
\label{eq:VlD}
\ee
The components of $V^\prime_D$ are dictated by the minimization conditions \eqref{eq:minD}
and the mass definitions of \eqref{eq:maD}.

\subsection{Vacuum of type $E$}
\label{sec:minE}

In an extremum of type $E$, according to table \ref{tab:vac}, only the singlet
$\chi_1$ acquires a vev:
\be
\langle \Phi\rangle_E\,=\,0\;\;\; , \;\;\; \langle \chi_1\rangle_E\,=\,w_E
\;\;\; , \;\;\; \langle \chi_2\rangle_E\,=\,0\, .
\ee
The stationarity condition of the potential is
\ba
\frac{\partial V}{\partial \chi_1}\,=\,0 &\Leftrightarrow & w_E
\left( \mu_2^2\,+\,\frac{1}{2}\,\lambda_2\,w_E^2\right)\,=\,0\, ,
\label{eq:minE}
\ea
and the neutral scalar squared mass matrix is given by
\be
\left[M^2\right]_E\,=\,\left(\begin{array}{ccc}
\mu_1^2\,+\,\frac{1}{2}\,\lambda_4\,w_E^2 & 0 & 0 \\
0 & \lambda_2\,w_E^2 & 0 \\
0 & 0 & \mu_3^2\,+\,\frac{1}{2}\,\lambda_6\,w_E^2 \\
\end{array}\right)
\ee
The three squared neutral masses at this stationary point are therefore given by
\be
m^2_{E1}\,=\,\mu_1^2\,+\,\frac{1}{2}\,\lambda_4\,w_E^2\;\;\; , \;\;\; m^2_{E2}\,=\,\lambda_2\,w_E^2
\;\;\; , \;\;\; m^2_{E3}\,=\,\mu_3^2\,+\,\frac{1}{2}\,\lambda_6\,w_E^2\,.
\label{eq:maE}
\ee
Electroweak symmetry is not broken, and therefore the fields $\varphi_1$, $\varphi_2$ and $\varphi_4$
all have squared masses equal to that of $\varphi_3$, meaning $m^2_{E1}$. An extremum of type $E$ is
therefore a minimum if and only if eq. \eqref{eq:minE} admits a solution and all squared masses in
eq. \eqref{eq:maE} are positive.

For later convenience, one defines the following two vectors:
\be
X_E\,=\,\langle X\rangle_E\,=\,\frac{1}{2}\,\left(\begin{array}{c} 0 \\ w_E^2 \\ 0 \end{array}\right)
\;\;\; , \;\;\;
V^\prime_E \,=\, A \,+\,B\,X_E \,=\, \left(\begin{array}{c} m^2_{E1} \\ 0 \\ m^2_{E3} \end{array}\right)\, .
\label{eq:VlE}
\ee
The components of $V^\prime_E$ are dictated by the minimization condition \eqref{eq:minE}
and the mass definitions of \eqref{eq:maE}.

\subsection{Vacuum of type $F$}
\label{sec:minF}

In an extremum of type $F$, according to table \ref{tab:vac}, only the singlet
$\chi_2$ acquires a vev, and the results are very similar to those obtained
for the previous case:
\be
\langle \Phi\rangle_F\,=\,0\;\;\; , \;\;\; \langle \chi_1\rangle_F\,=\,0
\;\;\; , \;\;\; \langle \chi_2\rangle_F\,=\,z_F\, .
\ee
The stationarity condition of the potential is
\ba
\frac{\partial V}{\partial \chi_2}\,=\,0 &\Leftrightarrow & z_F
\left( \mu_3^2\,+\,\frac{1}{2}\,\lambda_3\,z_F^2\right)\,=\,0\, .
\label{eq:minF}
\ea
The neutral scalar squared mass matrix is written as
\be
\left[M^2\right]_F\,=\,\left(\begin{array}{ccc}
\mu_1^2\,+\,\frac{1}{2}\,\lambda_5\,z_F^2 & 0 & 0 \\
0 & \mu_2^2\,+\,\frac{1}{2}\,\lambda_6\,z_F^2 & 0 \\
0 & 0 & \lambda_3\,z_F^2 \\
\end{array}\right)
\ee
and the squared neutral masses are then
\be
m^2_{F1}\,=\,\mu_1^2\,+\,\frac{1}{2}\,\lambda_5\,z_F^2\;\;\; , \;\;\;
m^2_{F2}\,=\,\mu_2^2\,+\,\frac{1}{2}\,\lambda_6\,z_F^2
\;\;\; , \;\;\; m^2_{F3}\,=\,\lambda_3\,z_F^2\,.
\label{eq:maF}
\ee
Electroweak symmetry is not broken, and thus the fields $\varphi_1$, $\varphi_2$ and $\varphi_4$
have squared masses equal to $m^2_{F1}$. An extremum of type $F$ is therefore a minimum if and only if
eq. \eqref{eq:minF} admits a solution and all squared masses in eq. \eqref{eq:maF} are positive.

For later convenience, one defines the following two vectors:
\be
X_F\,=\,\langle X\rangle_F\,=\,\frac{1}{2}\,\left(\begin{array}{c} 0 \\ 0 \\ z_F^2 \end{array}\right)
\;\;\; , \;\;\;
V^\prime_F \,=\, A \,+\,B\,X_F \,=\, \left(\begin{array}{c} m^2_{F1} \\ m^2_{F2} \\ 0 \end{array}\right)\, .
\label{eq:VlF}
\ee
The components of $V^\prime_F$ are dictated by the minimization condition \eqref{eq:minF}
and the mass definitions of \eqref{eq:maF}.

\subsection{Vacuum of type $G$}
\label{sec:minG}

In an extremum of type $G$, according to table \ref{tab:vac}, both singlets
acquires a vev:
\be
\langle \Phi\rangle_G\,=\,0\;\;\; , \;\;\; \langle \chi_1\rangle_G\,=\,w_G
\;\;\; , \;\;\; \langle \chi_2\rangle_G\,=\,z_G\, .
\ee
The stationarity condition of the potential are
\ba
\frac{\partial V}{\partial \chi_1}\,=\,0 &\Leftrightarrow & w_G
\left( \mu_2^2\,+\,\frac{1}{2}\,\lambda_2\,w_G^2\,+\,\frac{1}{2}\,\lambda_6\,z_G^2\right)\,=\,0\, . \nonumber \\
\frac{\partial V}{\partial \chi_2}\,=\,0 &\Leftrightarrow & z_G
\left( \mu_3^2\,+\,\frac{1}{2}\,\lambda_6\,w_G^2\,+\,\frac{1}{2}\,\lambda_3\,z_G^2\right)\,=\,0\, .
\label{eq:minG}
\ea
The neutral scalar squared mass matrix is then
\be
\left[M^2\right]_G\,=\,\left(\begin{array}{ccc}
\mu_1^2\,+\,\frac{1}{2}\,\lambda_4\,w_G^2\,+\,\frac{1}{2}\,\lambda_5\,z_G^2 & 0 & 0 \\
0 & \lambda_2\,w_G^2 & \lambda_6\,w_G\,z_G \\
0 & \lambda_6\,w_G\,z_G & \lambda_3\,z_G^2 \\
\end{array}\right)
\ee
and the squared neutral masses are then
\be
m^2_{G1}\,=\,\mu_1^2\,+\,\frac{1}{2}\,\lambda_4\,w_G^2\,+\,\frac{1}{2}\,\lambda_5\,z_G^2\;\;\; , \;\;\;
m^2_{G2,3}\,=\,\frac{1}{2}\left[ \lambda_2\,w_G^2\,+\,\lambda_3\,z_G^2\,\pm
\sqrt{\left(\lambda_2\,w_G^2\,-\,\lambda_3\,z_G^2\right)^2\,+\,4\,\lambda_6^2\,w_G^2\,z_G^2}\,\right]
\label{eq:maG}
\ee
The fields $\varphi_1$, $\varphi_2$ and $\varphi_4$
have squared masses equal to $m^2_{G1}$. An extremum of type $G$ is therefore a minimum if and only if
eqs. \eqref{eq:minG} admits a solution and all squared masses in eq. \eqref{eq:maG} are positive.

For later convenience, one defines the following two vectors:
\be
X_G\,=\,\langle X\rangle_G\,=\,\frac{1}{2}\,\left(\begin{array}{c} 0 \\ w_G^2 \\ z_G^2 \end{array}\right)
\;\;\; , \;\;\;
V^\prime_G \,=\, A \,+\,B\,X_G \,=\, \left(\begin{array}{c} m^2_{G1} \\ 0 \\ 0 \end{array}\right)\, .
\label{eq:VlG}
\ee
The components of $V^\prime_G$ are dictated by the minimization conditions \eqref{eq:minG}
and the mass definitions of \eqref{eq:maG}.

\section{Stability of minima}
\label{sec:sta}

Having established the conditions under which all types of minima can exist, a quick numeric
scan was performed, to verify that indeed all seven types of stationary points could be minima.
Indeed, for suitable choices of parameters, any of the seven types of stationary points can
be minima. In this section the possibility of more than one stationary point occurring in the
potential will be analysed.
As will be shown, that can indeed happen for many choices of parameters of the model. And it is
not guaranteed that, if one finds a minimum of a given type, it is the {\em global} minimum --
there are cases in which a deeper stationary point, of a different type, exists. In some cases,
the stability of a given minimum -- {\em i.e.} if that minimum exists, it is the global minimum
of the potential -- can be demonstrated analytically. If a minimum is found to be potentially
unstable -- {\em i.e.} a deeper stationary point, to which the system can eventually transit
via quantum tunneling -- a numerical study is necessary to determine the regions of parameter
space for which that may occur.

Since there are seven types of possible minima (check table \ref{tab:vac}), there are twenty--one
pairs of minima to compare. Only a few (typical) examples will be worked out in detail below,
illustrating the techniques that are used to compare the relative depths of the potential at
different extrema. The complete results will be shown in appendix~\ref{app:full}. In
appendix~\ref{app:U1} we briefly discuss the vacuum structure of the version of the model
which includes a global $U(1)$ symmetry.

\subsection{Extrema A {\em vs.} B}
\label{sec:ex1}

The starting point in the analysis in this section is the assumption that the potential
has two stationary points,
one of type $A$ -- meaning that eq.~\eqref{eq:minA} has a solution -- and another of type $B$ --
eqs.~\eqref{eq:minB} have a solution. These extrema are not required, {\em a priori}, to be minima.
Recall now the vectors defined in eqs.~\eqref{eq:VlA} and~\eqref{eq:VlB}, containing the
vevs of each extrema and information about the respective stationarity conditions:
\be
X_A\,=\,\frac{1}{2}\,\left(\begin{array}{c} v_A^2 \\ 0 \\ 0 \end{array}\right)
\;\; , \;\;
V^\prime_A \,=\, A \,+\,B\,X_A \,=\, \left(\begin{array}{c} 0 \\ m^2_{A2} \\ m^2_{A3} \end{array}\right)
\;\; , \;\;
X_B\,=\,\frac{1}{2}\,\left(\begin{array}{c} v_B^2 \\ w_B^2 \\ 0 \end{array}\right)
\;\; , \;\;
V^\prime_B \,=\, A \,+\,B\,X_B \,=\, \left(\begin{array}{c} 0 \\ 0 \\ m^2_{B3} \end{array}\right)\, .
\ee

The internal product of vectors $X_B$ and $V^\prime_A$ yields
\be
X_B^T\,V^\prime_A\,=\,\frac{1}{2}\,w_B^2\,m^2_{A2}
\label{eq:xbva}
\ee
but can also be expressed as
\be
X_B^T\,V^\prime_A\;=\;X_B^T\,\left(A \,+\,B\,X_A\right)\;=\;X_B^T\,A\,+\,X_B^T\,B\,X_A\,.
\label{eq:xbva2}
\ee

At this stage one sees, from eq.~\eqref{eq:Vmin}, that the quantity $X_B^T\,A$ is none other
than twice the value of the potential at the extremum $B$,
\be
X_B^T\,A\;=\; 2\,V_B\, ,
\ee
and as such, joining the results of eqs.~\eqref{eq:xbva} and~\eqref{eq:xbva2}, one obtains
\be
X_B^T\,B\,X_A \,=\, \frac{1}{2}\,w_B^2\,m^2_{A2} \,-\,2\,V_B\,.
\label{eq:xx1}
\ee
Performing similar operations upon the vectors $X_A$ and $V^\prime_B$, one obtains
\be
X_A^T\,V^\prime_B\,=\,0\,\Leftrightarrow\,X_A^T\,A\,+\,X_A^T\,B\,X_B\,=\,0\,.
\ee
The quantity $X_A^T\,A$ is twice the value of the potential at the extremum $A$, and as such
\be
X_A^T\,B\,X_B\,=\,-\,2\,V_A\,.
\label{eq:xx2}
\ee
Recalling that the matrix $B$ (defined in eq.~\eqref{eq:def}) is symmetric, the left-hand sides of
eqs.~\eqref{eq:xx1} and~\eqref{eq:xx2} are therefore equal and one finally obtains an expression
comparing the depth of the potential at the two extrema,

\be
V_B \,-\,V_A\,=\,\frac{1}{4}\,w_B^2\,m^2_{A2}\,.
\label{eq:vab1}
\ee

The significance of this equation is simple to understand: if $A$ is a minimum of the potential all of
its squared masses will be positive, and therefore $V_B \,-\,V_A\,>\,0$ --- the extremum $B$ lies
necessarily {\em above} the minimum $A$.

Moreover, the difference in the depths of the potential can also be expressed in terms of the masses
of the $B$ extremum. Recall that $v_A^2 \,=\, -\,2\,\mu_1^2/\lambda_1$ (eq.~\eqref{eq:minA}) and
therefore, from eq.~\eqref{eq:maA}, one has
\be
m^2_{A2}\,=\,\mu_2^2\,+\,\frac{1}{2}\,\lambda_4\,v_A^2\,=\,\mu_2^2\,-\,\frac{\lambda_4}{\lambda_1}\,\mu_1^2\,.
\label{eq:ma2}
\ee
Using the minimization conditions for the extremum $B$, eqs.~\eqref{eq:minB}, it can be found
that
\be
\mu_1^2\,=\,-\frac{1}{2}\left(\lambda_1\,v_B^2\,+\,\lambda_4\,w_B^2\right)\;\;\;,\;\;\;
\mu_2^2\,=\,-\frac{1}{2}\left(\lambda_4\,v_B^2\,+\,\lambda_2\,w_B^2\right)\,.
\ee
Therefore eq.~\eqref{eq:ma2} may be rewritten as
\be
m^2_{A2}\,=\,-\,\frac{\lambda_1\,\lambda_2\,-\,\lambda_4^2}{2\,\lambda_1}\,w_B^2\,=\,
-\,\frac{m^2_{B1}\,m^2_{B2}}{2\,\lambda_1\,v_B^2}\, ,
\ee
the last step in this deduction stemming from the definition of the squared masses in the
$B$ extremum, eq.~\eqref{eq:maB}. Introducing this result in eq.~\eqref{eq:vab1}, one
obtains

\be
V_B \,-\,V_A\,=\,\frac{1}{4}\,w_B^2\,m^2_{A2}\,\,=\,
-\,\frac{w_B^2}{8\,\lambda_1\,v_B^2}\,m^2_{B1}\,m^2_{B2}.
\label{eq:VAB}
\ee

Recall that one of the conditions that ensures the scalar potential is bounded from below
is $\lambda_1\,>\,0$ (eq.~\eqref{eq:bfb1}). This equation then implies that:
\begin{itemize}
\item {\em If $A$ is a minimum}, then $V_B \,-\,V_A\,>\,0$ which means that $m^2_{B1}\,m^2_{B2}\,<\,0$
 --- therefore, $A$ is deeper than $B$, and $B$ cannot be a minimum itself. In fact, it is
 necessarily a saddle point, since one of its squared masses is positive and the other negative.
\item {\em If $B$ is a minimum}, then $V_B \,-\,V_A\,<\,0$, which means that $m^2_{A2}\,<\,0$
 --- therefore $B$ is deeper than $A$, and $A$ cannot be a minimum itself. In fact, since
$m^2_{A1}\,=\,\lambda_1\,v_A^2$ and $\lambda_1\,>\,0$, the conclusion is that $A$ is necessarily a
saddle point.
\end{itemize}

Hence, if $A$ and $B$ extrema exist, and one of them is a minimum, the other is necessarily a
saddle point lying above it. Therefore, a minimum $A$ cannot ever tunnel to a deeper $B$ extremum,
and vice-versa.

\subsection{Extrema B {\em vs.} D}
\label{sec:ex2}

Suppose now that the potential has two extrema, of types of types $B$ and $D$.
This means that eqs.~\eqref{eq:minB} and
\eqref{eq:minD} have solutions, which may correspond, or not, to minima. Using the definitions of
eqs.~\eqref{eq:VlB} and~\eqref{eq:VlD}, it is possible to perform a set of calculations similar
to those of the last subsection:
\be
X_D^T\,V^\prime_B\,=\,\frac{1}{2}\,z_D^2\,m^2_{B3}
\label{eq:xdvb}
\ee
which may also be written as
\be
X_D^T\,V^\prime_B\;=\;X_D^T\,\left(A \,+\,B\,X_B\right)\;=\;X_D^T\,A\,+\,X_D^T\,B\,X_B\,.
\label{eq:xdvb2}
\ee
Using eq.~\eqref{eq:Vmin} again, the quantity $X_D^T\,A$ is twice the value of the potential
at the extremum $D$,
\be
X_D^T\,A\;=\; 2\,V_D\, ,
\ee
and hence
\be
X_D^T\,B\,X_B\,=\, \frac{1}{2}\,z_D^2\,m^2_{B3} \,-\,2\,V_D\,.
\ee
Similar operations are performed using $X_B$ and $V^\prime_D$, which are simplified
since $V^\prime_D \,=\,0$ (eq. \eqref{eq:VlD}), giving
\be
X_B^T\,V^\prime_D\,=\,0\,\Leftrightarrow\,X_B^T\,A\,+\,X_B^T\,B\,X_D\,=\,0
\,\Leftrightarrow\,X_B^T\,B\,X_D\,=\,-\,2\,V_B\,.
\label{eq:xbvld}
\ee
Comparing this expression with that of eq.~\eqref{eq:xdvb2}, one then obtains
\be
V_D \,-\,V_B\,=\,\frac{1}{4}\,z_B^2\,m^2_{B3}\,,
\label{eq:vbd1}
\ee
which tells us that if $B$ is a minimum, then it necessarily lies {\em deeper} than the
stationary point $D$.

Another useful expression for the depth difference in the potential is obtained
observing that
\be
V^\prime_D \,=\,A \,+\,B\,X_D\,=\,0\,\Leftrightarrow\,A\,=\,-\,B\,X_D\,.
\ee
Then it is possible to obtain
\be
V^\prime_B \,=\,A \,+\,B\,X_B\,\Leftrightarrow\,V^\prime_B \,=\,-\,B\,X_D \,+\,B\,X_B\,,
\ee
and therefore
\ba
B^{-1}V^\prime_B &=& -\,X_D \,+\,X_B \;\;\,\Leftrightarrow\,\nonumber \vspace{0.5cm}\\
\,\Leftrightarrow\,{V^\prime_B}^T B^{-1}V^\prime_B &=& - {V^\prime_B}^T \,X_D \,,
\ea
since, from eq.~\eqref{eq:VlB}, ${V^\prime_B}^T\,X_B = 0$. Comparing this equation
with \eqref{eq:xbvld} and \eqref{eq:vbd1}, it is easy to obtain
\be
V_D \,-\,V_B\,=\,\frac{1}{4}\,z_B^2\,m^2_{B3}\,=\,-\,\frac{1}{2}{V^\prime_B}^T B^{-1}V^\prime_B\,.
\label{eq:VBD}
\ee

Recall now eq.~\eqref{eq:maD}, which shows the mass matrix for the extremum $D$: it shows that
 for $D$ to be a minimum, the matrix $B$ -- and therefore also its inverse, $B^{-1}$ -- needs to
 have all of its eigenvalues be positive. Equation~\eqref{eq:VBD} then implies that:
\begin{itemize}
\item {\em If $B$ is a minimum}, then $V_D \,-\,V_B>0$ which means that
${V^\prime_B}^T B^{-1}V^\prime_B\,<\,0$ -- this implies that the matrix $B^{-1}$ cannot be positive-definite,
 and therefore neither can the matrix $B$. Accordingly, from eq.~\eqref{eq:maD} one concludes that at least one squared scalar mass is negative at the extremum $D$. However, at least one of those masses is positive, since all
 diagonal elements of $B$ are positive. Therefore, $B$ is deeper than $D$, and $D$ is necessarily a saddle point.
\item {\em If $D$ is a minimum}, all of its squared masses are positive and the matrix $B$ is positive definite.
Then $V_D \,-\,V_B<0$, which means that $m^2_{B3}\,<\,0$
 --- therefore $D$ is deeper than $B$, and $B$ cannot be a minimum itself. The bounded-from-below conditions
from eqs. \eqref{eq:bfb1} and \eqref{eq:bfb2} imply that ({\em cf.} eq. \eqref{eq:maB}) $m^2_{B1}>0$
and $m^2_{B2}>0$, so the conclusion is that $B$ is necessarily a saddle point.
\end{itemize}

This conclusion is valid for comparison of any type of extrema with extrema of type $D$ -- one
obtains always expressions similar to eq.~\eqref{eq:VBD}, which imply that when $D$ is a minimum, it is
always the deepest one, and the other extremum is a saddle point (see appendix~\ref{app:full}). This
already permits a conclusion -- if the potential has a minimum of type $D$, that is the global
minimum of the model, and there are no other minima, only eventually saddle points (or a maximum at the
origin).

\subsection{Extrema $B$ {\em vs.} $G$}
\label{sec:ex3}

In the previous subsections, comparison between certain pairs of extrema showed
that when one of
them was a minimum, the other was necessarily a saddle point and lay above the minimum.
The minimum could not therefore tunnel to the other extremum, and its stability was assured.
As will now be shown, that is not the case for other possible pairs of extrema.

The procedure is the same that was outlined in previous subsections: one starts with the
vectors defined in eqs.~\eqref{eq:VlB} and~\eqref{eq:VlG}, and performs internal products
between some of them:
\begin{align}
 X_B^T\,V^\prime_G \quad=\quad  X_B^T\,A\,+\, X_B^T\,B\,X_G \quad&\Leftrightarrow   \quad
\frac{1}{2}\,v_B^2\,m^2_{G1} \quad=\quad 2\,V_B \,+\, X_B^T\,B\,X_G\, , \nonumber \\
 X_G^T\,V^\prime_B \quad=\quad  X_G^T\,A\,+\, X_G^T\,B\,X_B  \quad& \Leftrightarrow  \quad
\frac{1}{2}\,z_G^2\,m^2_{B3} \quad=\quad 2\,V_G \,+\, X_G^T\,B\,X_B\, .
\end{align}
Subtracting both equations, one then obtains
\be
V_G \,-\,V_B\,=\,\frac{1}{4}\,\left(z_G^2\,m^2_{B3}\,-\,v_B^2\,m^2_{G1}\right) \,.
\label{eq:VBG}
\ee

The right-hand side of this equation has no definite sign -- a numerical scan shows that,
depending on the parameters of the potential, either $B$ or $G$ can be deeper than the other,
and inclusively, they can {\em both} be minima, simultaneously. This is therefore a
qualitatively different situation from that found on the previous subsections. In fact,
eq.~\eqref{eq:VBG} implies that a minimum of type, say, $B$, if it exists, is not safe
against tunneling to a deeper $G$ minimum. To avoid this situation, one may impose cuts
on the parameters of the model to ensure that any extremum $G$, if they exist, lie
{\em above} $B$. This point will be addressed in section \ref{sec:con}.

In some cases the expressions for the depth difference can be further simplified,
putting the dependence on the parameters in sharp evidence. For instance, following
the steps outlined in this subsection for the case of two extrema $B$ and $C$, one
can obtain
\ba
V_C \,-\,V_B &=& \frac{1}{4}\,\left(z_C^2\,m^2_{B3}\,-\,w_B^2\,m^2_{C2}\right) \nonumber \\
  &=& \frac{1}{2}\,\left(\frac{\mu_2^4}{\lambda_2}\,-\,\frac{\mu_3^4}{\lambda_3}\right)\,.
\ea
This last expression makes it all the more obvious that specific choices of the parameters of the
potential may yield either of the extrema as the deepest. Again for this case, a numerical
scan of the potential shows that both extrema can be simultaneouly minima.

\subsection{Extrema $C$ {\em vs.} $E$}
\label{sec:ex4}

As a final example, consider two simultaneous extrema of the potential, $C$ and $E$. Using
the definitions of eqs. \eqref{eq:VlC} and \eqref{eq:VlE}, and repeating the procedure of
the previous subsection, one finds
\begin{align}
 X_E^T\,V^\prime_C \quad=\quad  X_E^T\,A\,+\, X_E^T\,B\,X_C \quad&\Leftrightarrow   \quad
\frac{1}{2}\,w_E^2\,m^2_{C2} \quad=\quad 2\,V_E \,+\, X_E^T\,B\,X_C\, , \nonumber \\
 X_C^T\,V^\prime_E \quad=\quad  X_C^T\,A\,+\, X_C^T\,B\,X_E  \quad& \Leftrightarrow  \quad
\frac{1}{2}\,\left(v_C^2\,m^2_{E1}\,+\,z_C^2\,m^2_{E3}\right)\quad=\quad 2\,V_C \,+\, X_C^T\,B\,X_E\, .
\end{align}
And thus, the difference in depths of the potential is given by
\be
V_E \,-\,V_C\,=\,\frac{1}{4}\,\left(w_E^2\,m^2_{C2}\,-\,v_C^2\,m^2_{E1}\,-\,z_C^2\,m^2_{E3}\right) \,.
\label{eq:VCE}
\ee
Again, this expression has no definite sign -- each of the extrema may be a minimum,
one is not privileged relative to the other, and either can be deeper than the other,
depending on specific values for the parameters in the potential.

\subsection{Global results}

The four examples shown in previous subsections are typical of the expressions obtained
for all 21 possible comparisons between pairs of extrema, and the deductions made to find
those expressions. The full expressions for all differences in depth of the potential
for pairs of extrema are shown in appendix \ref{app:full}. What overall conclusions may
one deduce from those results?

The two first examples (sections \ref{sec:ex1} and \ref{sec:ex2}) were of pairs of minima
where {\em Stability} was achieved: if one of the extrema was a minimum, the other was a saddle
point, guaranteed to be lying above the minimum.

On the other hand, the two following examples (sections \ref{sec:ex3} and \ref{sec:ex4}) were
of pairs of minima where an {\em Undefined} situation occurred: both extrema could simultaneously
be minima (as verified in numerical scans of the model), and neither was privileged compared
to the other -- depending on the particular choice of parameters made, either extrema could be
the deepest minimum.

These results are summarized in table \ref{tab:comp}. Some conclusions can immediately be drawn:
\begin{table*}[ht!]
\caption{\em Stability of pairs of extrema in the potential. For a given pair of extrema, ``Stability" means that if one of them is a minimum, the other is necessarily a saddle point lying above the minimum. A pair of ``Undefined" extrema
(marked below with ``$\times$")
means that both of them can be simultaneously minima, and neither is guaranteed to be the deepest one, depending on the
choice of parameters.}
\begin{tabular}{c|c|c|c|c|c|c|c}
\hline
\hline
Extrema & $A$ & $B$ & $C$ & $D$ & $E$ & $F$ & $G$ \\
\hline
 & & & & & & & \\
$A$ & --- & \;\;Stability\;\; & \;\;Stability\;\; & \;\;Stability\;\; & $\times$ & $\times$ & $\times$ \\
 & & & & & & & \\
 \hline
  & & & & & & & \\
$B$ & \;\;Stability\;\; & --- & $\times$ & Stability & Stability & $\times$  & $\times$ \\
 & & & & & & & \\
 \hline
  & & & & & & & \\
$C$ & Stability & $\times$ & --- & Stability & $\times$ & Stability & $\times$ \\
 & & & & & & & \\
 \hline
  & & & & & & & \\
$D$ & Stability & Stability & Stability & --- & \;\;Stability\;\; & \;\;Stability\;\; & \;\;Stability\;\; \\
 & & & & & & & \\
 \hline
  & & & & & & & \\
$E$ & $\times$ & Stability & $\times$ & Stability & --- & $\times$ & Stability \\
 & & & & & & & \\
 \hline
  & & & & & & & \\
$F$ & $\times$ & $\times$ & Stability & Stability & $\times$ & --- & Stability \\
 & & & & & & & \\
 \hline
  & & & & & & & \\
$G$ & $\times$ & $\times$ & $\times$ & Stability & Stability & Stability & --- \\
 & & & & & & & \\
\hline
\hline
\end{tabular}
\label{tab:comp}
\end{table*}

\begin{itemize}
\item The extremum $D$ -- where all fields acquire vevs -- is qualitatively different from
the remaining ones. If $D$ is a minimum, it is the global minimum of the model, and all other
extrema are saddle points. A perfectly stable vacuum, then, is the one that breaks all
symmetries of the model.
\item All extrema other than $D$ are stable against three other extrema, and have undefined
behaviour against three others. Their stability is therefore not guaranteed, and the
possibility arises to impose constraints on the parameter space of the model to ensure that a
given minimum is the global one.
\item A minimum of type $A$ is stable against tunneling for extrema $B$ and $C$, which could
be interpreted as extra evidence those two extrema are altogether physically equivalent,
and perhaps merely related via basis changes. However, notice that an extremum of type $B$ is
{\em not} stable with respect to an extremum $C$. Therefore, it is indeed necessary to consider
all possible types of extrema of the potential.
\item A numerical study of the several possibilities has shown that it is possible to find
values of the parameters of the potential such that, when ``stability" is not ensured
in table \ref{tab:comp}, those pairs of extrema are simultaneously minima.
\end{itemize}

\section{Constraints on the parameter space}
\label{sec:con}

The fact that (other than $D$) no minimum of the potential is guaranteed to be stable raises
the possibility of imposing constraints on the parameter space, to ensure that that
minimum is the global one. As was already explained, the only physically acceptable minima
are of types $A$ to $D$ -- in the remaining ones no electroweak breaking occurs and the
bosons $W$ and $Z$, as well as all fermions, would be massless. Minimum $A$ is SM-like, with
two neutral scalars that do not couple to fermions or gauge bosons, thus the model's scalar
spectrum includes a SM-like scalar and two dark matter candidates. Minima $B$ and $C$
present deviations from SM behaviour - two neutral scalars mix, the third one remaining
unmixed. Finally, in minimum $D$ all three neutral scalars mix, and there are no ``inert" scalars
that could naturally play the role of a dark matter candidate.

In what follows the phenomenology of a minimum of type
$B$ will be analysed~\footnote{A minimum of type $C$
has the same phenomenology as type $B$'s, so it would be an equally acceptable choice.},
since that model
seems the most interesting: the mixing of two scalars means that we can have a SM-like scalar,
but with deviations from SM expectations which might be testable at LHC; and the model also
includes a dark matter candidate. By contrast, a minimum of type $A$ would have a neutral
scalar virtually identical to the SM's, and the model becomes less interesting from a LHC
point of view. The $D$ minimum has interesting LHC phenomenology but no dark matter candidates,
which are a good motivation to expanding the scalar sector of the SM.

In studying the phenomenology of model $B$, there are some basic demands that must be met:
\begin{itemize}
\item The potential should be bounded from below, {\em i.e.} its quartic couplings must
obey the constraints of eqs. \eqref{eq:bfb1}--\eqref{eq:bfb2}.
\item So that unitarity and perturbativity are satisfied, all quartic couplings should be
suitably small. An upper bound, $|\lambda_i| < 10$, was imposed.
\item The model should include a SM-like neutral scalar with mass $m_h = 125$ GeV.
\item Another neutral state $H$, which couples to fermions and gauge bosons, is present
and heavier than $h$: $m_h < m_H < 1000$ GeV. The upper limit is somewhat
arbitrary, but larger values of $m_H$ endanger the perturbativity of the
model~\footnote{The possibility that another scalar, lighter
than 125 GeV, could have escaped detection thus far, has been considered \cite{Ferreira:2012my}, but
it corresponds to such a small region of parameter space that it is tantamount to fine-tuning.
Unless, of course, that state is an inert scalar and thus a dark matter candidate, which
is the case of the third neutral scalar of the current model.}.
\item The model's third neutral scalar is an inert scalar -- without interactions to gauge
bosons or fermions -- and thus a perfect dark matter candidate, $h_3$. We take $20 < m_{h_3} <
1000$ GeV.
\end{itemize}

For a minimum of type $B$, the neutral fields $\varphi_3$ and $\chi_1$ mix among themselves,
as one sees from the mass matrix~\eqref{eq:maB}. The mass eigenstates $h$ and $H$ are therefore
obtained from the diagonalization of~\eqref{eq:maB}, with a mixing angle $\alpha$ defined by
\be
h\,=\, \sin\alpha\,\varphi_3\,+\,\cos\alpha\,\chi_1\;\;\; , \;\;\;
H\,=\, -\cos\alpha\,\varphi_3\,+\,\sin\alpha\,\chi_1\, .
\label{eq:diag}
\ee
Recall from section \ref{sec:minB} that the doublet $\Phi$ acquires a vev $v_B/\sqrt{2}$
and the singlet $\chi_1$ a vev $w_B$. Since only the doublet gives mass to fermions
and the electroweak gauge bosons, it is mandatory that $v_B = 246$ GeV. As for the
singlet's vev, we choose $-1000 < w_B < 1000$ GeV. In terms of physical masses, vevs and
the diagonalization angle $\alpha$, expressions for three of the quartic couplings may
be obtained,
\be
\lambda_1\,=\,\frac{\sin^2\alpha\,m^2_h\,+\,\cos^2\alpha\,m^2_H}{v_B^2}\;\;\; , \;\;\;
\lambda_2\,=\,\frac{\cos^2\alpha\,m^2_h\,+\,\sin^2\alpha\,m^2_H}{w_B^2}\;\;\; , \;\;\;
\lambda_4\,=\,\frac{(m^2_h\,-\,m^2_H)\sin\alpha\cos\alpha}{v_B w_B}\, .
\label{eq:l14}
\ee
Inverting eq. \eqref{eq:diag} we obtain $\varphi_3 = \sin\alpha\,h\,-\,\cos\alpha\,H$.
Since only $\varphi_3$ couples to the fermions and electroweak gauge bosons, this model
has a very simple set of rules concerning the scalars' interactions:
\begin{itemize}
\item The couplings of the lightest scalar $h$ to fermions and gauge bosons are identical to
those of the SM Higgs boson, multiplied by a factor of $\sin\alpha$.
\item The couplings of the heavier scalar $H$ to fermions and gauge bosons are identical to
those of the SM Higgs boson, multiplied by a factor of $-\cos\alpha$.
\end{itemize}

The current status of the LHC experiments points to a scalar of mass 125 GeV having
interactions almost identical to those expected for the SM Higgs boson, with experimental
uncertainties of at most $\sim$ 10\% being measured in its couplings to gauge
bosons and fermions \cite{Khachatryan:2016vau}.
Therefore, taking $0.9 \,\leq \sin\alpha \leq \,1$,
the current experimental results will be adequately complied.

With the restrictions detailed above, a scan over the model's parameter space was
performed, generating over $10^7$ different combinations of free parameters --
having fixed $m_h$ and $v_B$, the model is left with seven free parameters, which
were taken to be $w_B$, $\sin\alpha$ ($\cos\alpha$ was allowed to have both signs),
$m_H$, $m_{h_3}$, and $\lambda_{3,5,6}$. Through eq. \eqref{eq:l14} the remaining quartic
couplings are determined; and the quadratic parameters of the potential \eqref{eq:Vr}
are determined from the minimisation conditions \eqref{eq:minB} ($\mu_1^2$ and
$\mu_2^2$) and from the definition of the third scalar mass from eq. \eqref{eq:maB}
($\mu_3^2$).

With all the parameters of the potential specified, and generated in such a way as to ensure
that it possesses a minimum of type $B$~\footnote{This is achieved by simultaneously satisfying
the extremum conditions of eqs.~\eqref{eq:minB} and using scalar masses as input parameters.}
is a simple matter to verify whether other, deeper, extrema are also present. From
table \ref{tab:comp}, recall that a minimum of type $B$ can only have deeper extrema of
types $C$, $F$ and $G$. Thus, for each set of parameters, it is verified whether
eqs.~\eqref{eq:minC}, \eqref{eq:minF} and \eqref{eq:minG} have non-trivial solutions --
if they do, the vevs corresponding to each of those extrema are obtained and the value
of the potential therein is computed, and compared with $V_B$ -- the value of the potential
at the minimum $B$.

\begin{figure}
\includegraphics[height=6cm,angle=0]{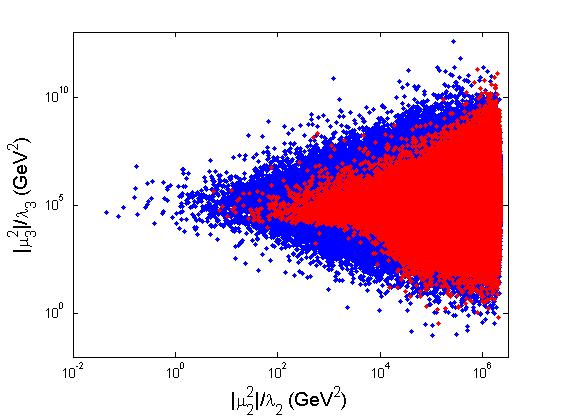}
\includegraphics[height=6cm,angle=0]{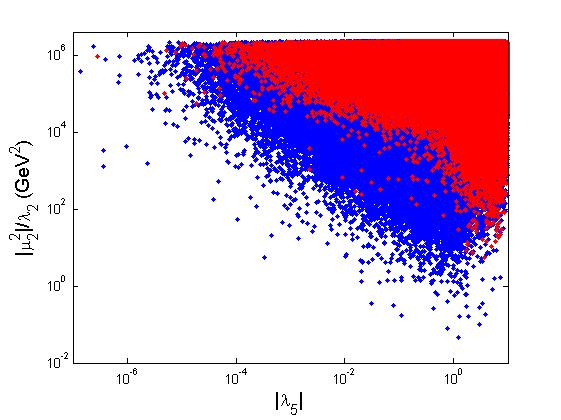}
\caption{Scatter plots of $\mu_2^2/\lambda_2$ {\em versus} $\mu_3^2/\lambda_3$ (left) and
$\lambda_5$ {\em versus} $\mu_2^2/\lambda_2$ (right). In red (grey), are points for which
$V_B > V_{C,F,G}$, with $V_X$ the value of the potential at extremum $X$. In blue
(dark) points for which $V_B < V_{C,F,G}$ -- and therefore points in which
$B$ is the global minimum.}
\label{fig:dV}
\end{figure}
The results of this procedure are shown in fig. \ref{fig:dV}: in the left plot,
the variable $\mu_2^2/\lambda_2$ is plotted against $\mu_3^2/\lambda_3$; in the right
plot,  the $\lambda_5$ quartic coupling is plotted against $\mu_2^2/\lambda_2$. The red (grey)
region corresponds to points of the parameter space (9-parameter combinations) for which
$B$ is {\em not} the global minimum of the potential -- a deeper extremum, of types $C$,
$F$ or $G$ exists, and the possibility of tunneling to it exists. In blue (dark) are the points
for which $B$ is indeed the global minimum of the potential, its stability therefore
ensured. The reader should exercise caution in interpreting these figures: though the
red (grey) regions seem dense, they in fact are not. There are plenty of acceptable
points in the middle of those regions, a consequence of the high dimensionality of the
parameter space. So for instance, though the region with $\mu_2^2/\lambda_2 \simeq 10^5$ GeV$^2$
and $\mu_3^2/\lambda_3 \simeq 10^5$ GeV$^2$ seems excluded in the first plot of fig. \ref{fig:dV},
this is not so: there are many points for which $B$ is the acceptable minimum and which
fill that region of parameter space.

Fig. \ref{fig:dV} shows that the regions of parameter space for which $B$ is not
the global minimum are not negligible -- in quantitative terms, the numerical
simulation herein performed involved a total of $10^7$ points, and of those about
11\% were found for which $B$ is not the global minimum~\footnote{The generation of points for
the parameter space was as random and dense as possible, but still, these 11\% do not have any
physical meaning. The number is merely an indication of the size of the excluded region.}.
It would be interesting if imposing a bound of absolute stability on the $B$ minimum would translate
into constraints on  physical parameters with easy-to-grasp meaning -- for instance,
a bound on the mass of $H$, on $\sin\alpha$ or the mass of the dark matter candidate.
Unfortunately, the absolute stability bound has no direct, visible, bearing on such
variables~\footnote{The same is true for other bounds imposed on the parameters of
the potential -- the bounded-from-below conditions of eqs.
\eqref{eq:bfb1}--\eqref{eq:bfb3}, for instance, do not lead to direct bounds on
parameters such as $m_H$, $\sin\alpha$ or $m_{h_3}$.}.

Finally, notice that the results shown in this section are obviously entirely analogous
to those one would find for a minimum of type $C$ -- because vacua $B$ and $C$ are
physically equivalent, since they break the same symmetries (although, it should be
re-emphasised, they can coexist with one another, albeit with different values, for
instance, of masses and vevs).

\section{Discussion and conclusions}
\label{sec:fin}

The work presented in this paper has demonstrated that in a model with a scalar doublet
and a complex singlet there are seven different types of possible vacua. And also
that, with a single exception -- a minimum of type $D$, where all neutral fields
acquire vevs -- the occurrence of a minimum in the potential does not guarantee that
it be the global minimum. In fact, with the exception of $D$, all possible minima can
have one or more of three different types of minima lower than them, so that their stability
is not guaranteed. The detailed numerical study presented in section \ref{sec:con} has
shown that the region of parameter space where deeper minima coexist with a minimum of
type $B$ -- which is arguably the most physically interesting type of minimum, since
it breaks electroweak symmetry, has a richer scalar sector than the SM's that can potentially
be probed at the LHC and boasts a dark matter candidate -- is small, and has no
substantial phenomenological implications for the model. But nevertheless, the possibility
of unstable minima is present in the model, and ought at least to be considered -- it is a very
simple matter to apply the formulae supplied in this work to verify whether, for a given
choice of parameters of the potential, the minimum under consideration is global,
or rather a metastable one. That comparison is made even simpler by the fact that the
minimisation conditions which determine the vevs at each extremum are analytically solvable.

All the discussion so far has only considered the possibility of different extrema coexisting,
and a given minimum not being global -- so that it could, in principle, tunnel to a lower
one. This could mean that the minimum the Universe is currently occupying -- with electroweak
symmetry broken, $v = 246$ GeV, all particles with their known masses -- could be unstable
and tunnel to the global minimum of the model, where electroweak symmetry could even be
unbroken, or the value of the doublet's vev be different, which would change the masses of
all elementary particles. Both situations would be catastrophic and should be avoided at
all costs. Of course, this situation is only problematic if the tunneling time between
minima is inferior to the age of the Universe. The calculation of the tunneling time between
two minima was first presented in the classic works by Coleman {\em et al}
\cite{Coleman:1977py,Callan:1977pt}, for the case of a single real scalar field. The cases
discussed here involve three scalars, so the problem becomes considerably more complex,
since it is not obvious what is the best bounce trajectory in the multi-dimensional Higgs space
of the potential. In simple terms \cite{coleman_book,Rubakov:2002fi,Adams:1993zs}, if
$\delta$ is the height of the ``potential barrier" between
two minima and $\epsilon$ their relative depth, the value of the quantity $|\delta/\epsilon|$
provides a quick estimate of the dangers of tunneling: if $|\delta/\epsilon|  < 1$, the
lifetime of the local minimum tends to be quite small, and such minima would in principle
become unstable during the time elapsed since the Big Bang; on the other hand, if
$|\delta/\epsilon| \gtrsim 1$, the false vacuum tends to have a lifetime larger than the age
of the Universe, and as such it may be regarded as safe. Looking at the numerical scan
presented in section \ref{sec:con}, while comparing two extrema, $B$ and $X$ (with $X$ = $C$, $F$, $G$),
$\delta$ was taken as $\delta = |max(V_B\, ,\,V_X)|$ (assuming thus that the maximum at the origin
corresponds to the barrier between both extrema) and $\epsilon$, $\epsilon = |V_B \,-\,V_X|$.
It is then found that for about half of the points for which $B$ is a local minimum, {\em i.e}
$V_B > V_X$, for some $X$, one has $|\delta/\epsilon|  < 1$, and thus tunneling could have occurred
within the Universe's lifespan.

There are however shortcomings in this lifetime estimate. To begin with, these calculations
use the so-called ``thin-wall" approximation, in which the border between the regions of the
universe lying in different vacua is considered to have no thickness. This approximation breaks
down if $|\delta/\epsilon| \leq 0.1$ -- and for most of the local $B$ minima found, $\delta/\epsilon$
is much smaller than this upper bound, which makes this estimate problematic.
Maybe more serious, however, the estimates that were shown, while used often in the literature,
concern a potential with a single scalar field, whereas the model under discussion has three.
The ``bounce trajectory" between two minima is therefore not obvious, the possibility of
trajectories which avoid intermediate extrema a possibility which needs to be considered
\cite{Rubakov:2002fi} -- which means that there may be ``easier" paths between minima than
the ones that were considered above, and which would greatly reduce the tunneling time. Thus,
for plenty of local minima that seem to be safe against tunneling, a more exhaustive calculation
might indeed turn out to reveal a tunneling time inferior to the age of the Universe.
This was also the case for similar concerns within the 2HDM, wherein the electroweak vacuum
turns out, for some choices of parameters, to be a local minimum \cite{Barroso:2013awa}.

The occurrence of metastable minima is also a feature of the SM, of course -- in
fact, the observed Higgs mass of 125 GeV suggests that the SM electroweak breaking
vacuum is metastable, once loop corrections are taken into account \cite{Degrassi:2012ry}.
But within the SM metastability is a consequence of quantum corrections to the potential
\cite{Cabibbo:1979ay,Lindner:1985uk,Sher:1988mj,Altarelli:1994rb,Casas:1994qy,Casas:1996aq,
Isidori:2001bm} -- in the singlet-doublet model herein discussed, metastability arises
at tree-level already. The same was verified for the 2HDM potential
\cite{Barroso:2012mj,Barroso:2013awa}. It is to be expected that the results found here could be
substantially altered once loop corrections are taken into account -- indeed, in the SM
one can pass from a single minimum at tree-level to a metastable one at one-loop.
Therefore, it is certainly possible that quantum corrections enlarge significantly the region
of parameter space for which a local $B$ minimum, for instance, is unstable. It should be noted that,
within the context of inert minima in the 2HDM, it was found in \cite{Ferreira:2015pfi} that
the tree-level  constraints for the coexistence of two minima were considerably relaxed
once one-loop  corrections were taken into account. Of course, another possibility is that the
one-loop analysis - involving the running of all couplings of the model and the one-loop
potential -- reveals that tree-level instability is cured at the one-loop level. In fact,
in \cite{Ferreira:2015pfi}, a small number of local minima at tree-level were promoted to
global ones at one-loop.

On a more fundamental note, the results presented in this work reveal a rich vacuum structure
of the singlet-doublet model, one in which minima which break different symmetries are
allowed to coexist with one another. This is in stark contrast with the 2HDM situation: in
\cite{Ivanov:2006yq,Ivanov:2007de} it was proven that minima of different natures --
in which different symmetries were broken -- could not coexist. For the 2HDM only electroweak
breaking minima can (and do, for certain regions of the model's parameter space) exist
simultaneously, a fact that was then exploited in refs.
\cite{Barroso:2012mj,Barroso:2013awa,Ferreira:2015pfi}. The results of the analysis herein
presented show that the ``theorem" which was valid for the 2HDM  -- minima of different natures
cannot coexist -- is not valid in other models.

Finally, a word on soft breaking terms: it is very common to consider potentials for the
complex singlet-doublet model where the global symmetries imposed (discrete $Z_2$'s or global
$U(1)$) are broken by means of soft breaking terms in the potential. Meaning, terms involving
powers of the singlets smaller than four -- these terms do not spoil the renormalizability of
the model, and enlarge the allowed parameter space. Since the singlets carry no gauge numbers,
even linear soft breaking terms are allowed. Those terms linear in the singlet fields, though
unusual, play an interesting role, preventing the occurrence of cosmological
domain walls \cite{Barger:2008jx}. Other soft breaking terms can also, for instance, provide a mass
for massless scalars originated from the breaking of the global $U(1)$ symmetry. In terms of
the real component formalism used in this work, soft breaking terms would introduce terms in
the potential with odd powers of $\chi_1$ and $\chi_2$-- terms such as $\chi_1^3$, $\chi_1 \chi_2^2$,
$\chi_1^3 \chi_2$, etc. The number of possible vacua would remain unchanged, but the analysis
would become much more complex. In particular, the bilinear formalism used in section
\ref{sec:min} would no longer be viable, and the analytical formulae obtained for the
comparison of the values of the potential at each extremum would be substantially changed --
in fact, most of the minimization conditions would cease to have analytical solutions, and only
numerical comparisons between extrema would be possible. It is certain that the picture
exposed in this paper -- of stability of some vacua regarding others -- would be completely
altered, and one can expect that no minimum would be guaranteed to be safe against
tunneling against another one.

In summary, the vacuum structure of the complex singlet-doublet model is rich and complex --
possible minima of seven different types are possible. The version of the model chosen here
includes two discrete symmetries imposed upon its singlet fields, simplifying considerably
the form of the potential, but still leaving it with a nine-dimensional parameter space.
It was shown that, for specific values
of the parameters of the model, different minima can coexist in the potential, and, with a
single exception, the stability of a given minimum is not guaranteed. The one minimum which,
if it exists, is guaranteed to be global, is the one dubbed here as type $D$ -- a vacuum in
which all fields acquire a vev and thus all symmetries (electroweak and discrete) are broken.
For a minimum of the remaining six types, they are guaranteed to be deeper than three other
stationary points (which, if they exist as solutions of the minimisation equations of the potential,
are forced to be saddle points). But they may, or may not, lie above one or more three other extrema,
which may, or may not, also be minima. A thorough numerical scan of one of the models was then
undertaken. The model corresponds to a minimum of type $B$, where electroweak symmetry, and one of
the discrete ones, is broken by vevs acquired by two of the fields (the doublet and one of the
singlets). The parameters were chosen such that the model boasted a SM-like Higgs scalar of
mass 125 GeV, to conform with LHC observations. It was found that a small percentage of
points in the parameter space produced local $B$ minima, which could tunnel to
deeper, global minima.

\acknowledgments{Many thanks to Lu\'\i s Lavoura, Marco Sampaio and Rui Santos
for discussions concerning the possibility of CP breaking in the model. I am 
further indebted to Lu\'\i s Lavoura for a careful and critical read of the 
manuscript.
}

\appendix

\section{Full results on minima depth comparison}
\label{app:full}

In this appendix the formulae for the comparison of the value of the potential
at all possible pairs of extrema are presented. The demonstration of these results
is not shown -- it follows entirely the type of calculations shown in
section \ref{sec:sta}, which can be trivially adapted for the cases shown
below. For completion, we repeat the results from subsections
\ref{sec:ex1}--\ref{sec:ex4}. Some, but not all, of the formulae can be simplified
and provide expressions in terms of the potential's parameters instead of in terms
of vevs or masses.

\begin{itemize}

\item For a pair of extrema $A$ and $B$,
\be
V_B \,-\,V_A\,=\,\frac{1}{4}\,w_B^2\,m^2_{A2}\,\,=\,
-\,\frac{w_B^2}{8\,\lambda_1\,v_B^2}\,m^2_{B1}\,m^2_{B2}.
\ee
Thus if one of them is a minimum it is the deeper extremum, and
the other a saddle point. Hence the minimum is stable.
\item For a pair of extrema $A$ and $C$,
\be
V_C \,-\,V_A\,=\,\frac{1}{4}\,z_C^2\,m^2_{A3}\,\,=\,
-\,\frac{z_C^2}{8\,\lambda_1\,v_C^2}\,m^2_{C1}\,m^2_{C3}.
\ee
Thus if one of them is a minimum it is the deeper extremum, and
the other a saddle point. Hence the minimum is stable.
\item For a pair of extrema $A$ and $D$,
\be
V_D \,-\,V_A\,=\,\frac{1}{4}\,\left(w_D^2\,m^2_{A2}\,+\,z_D^2\,m^2_{A3}\right)\,=\,
-\,\frac{1}{2}{V^\prime_A}^T B^{-1}V^\prime_A.
\ee
Thus if one of them is a minimum it is the deeper extremum, and
the other a saddle point. Hence the minimum is stable.
\item For a pair of extrema $A$ and $E$,
\be
V_E \,-\,V_A\,=\,\frac{1}{4}\,\left(w_E^2\,m^2_{A2}\,-\,v_A^2\,m^2_{E1}\right)\,=\,
\frac{1}{2}\,\left(\frac{\mu_1^4}{\lambda_1}\,-\,\frac{\mu_2^4}{\lambda_2}\right).
\ee
Both extrema can be simultaneously minima, and none of them is necessarily deeper
than the other. Stability of either extrema is thus not ensured.
\item For a pair of extrema $A$ and $F$,
\be
V_F \,-\,V_A\,=\,\frac{1}{4}\,\left(z_F^2\,m^2_{A3}\,-\,v_A^2\,m^2_{F1}\right)\,=\,
\frac{1}{2}\,\left(\frac{\mu_1^4}{\lambda_1}\,-\,\frac{\mu_3^4}{\lambda_3}\right).
\ee
Both extrema can be simultaneously minima, and none of them is necessarily deeper
than the other. Stability of either extrema is thus not ensured.
\item For a pair of extrema $A$ and $G$,
\be
V_G \,-\,V_A\,=\,\frac{1}{4}\,\left(w_G^2\,m^2_{A2}\,+\,z_G^2\,m^2_{A3}\,-\,v_A^2\,m^2_{G1}\right).
\ee
Both extrema can be simultaneously minima, and none of them is necessarily deeper
than the other. Stability of either extrema is thus not ensured.
\item For a pair of extrema $B$ and $C$,
\be
V_C \,-\,V_B\,=\,\frac{1}{4}\,\left(z_C^2\,m^2_{B3}\,-\,w_B^2\,m^2_{C2}\right)\,=\,
\frac{1}{2}\,\left(\frac{\mu_2^4}{\lambda_2}\,-\,\frac{\mu_3^4}{\lambda_3}\right).
\ee
Both extrema can be simultaneously minima, and none of them is necessarily deeper
than the other. Stability of either extrema is thus not ensured.
\item For a pair of extrema $B$ and $D$,
\be
V_D \,-\,V_B\,=\,\,=\,\frac{1}{4}\,z_D^2\,m^2_{B3}\,=\,
-\,\frac{1}{2}{V^\prime_B}^T B^{-1}V^\prime_B\,.
\ee
Thus if one of them is a minimum it is the deeper extremum, and
the other a saddle point. Hence the minimum is stable.
\item For a pair of extrema $B$ and $E$,
\be
V_E \,-\,V_B\,=\,-\,\frac{1}{4}\,v_B^2\,m^2_{E1}\,\,=\,
\frac{v_B^2}{8\,\lambda_2\,w_B^2}\,m^2_{B1}\,m^2_{B2}.
\ee
Thus if one of them is a minimum it is the deeper extremum, and
the other a saddle point. Hence the minimum is stable.
\item For a pair of extrema $B$ and $F$,
\be
V_F \,-\,V_B\,=\,\frac{1}{4}\,\left(z_F^2\,m^2_{B3}\,-\,v_B^2\,m^2_{F1}\,-\,w_B^2\,m^2_{F2}\right).
\ee
Both extrema can be simultaneously minima, and none of them is necessarily deeper
than the other. Stability of either extrema is thus not ensured.
\item For a pair of extrema $B$ and $G$,
\be
V_G \,-\,V_B\,=\,\frac{1}{4}\,\left(z_G^2\,m^2_{B3}\,-\,v_B^2\,m^2_{G1}\right).
\ee
Both extrema can be simultaneously minima, and none of them is necessarily deeper
than the other. Stability of either extrema is thus not ensured.
\item For a pair of extrema $C$ and $D$,
\be
V_D \,-\,V_C\,=\,\,=\,\frac{1}{4}\,w_D^2\,m^2_{C2}\,=\,
-\,\frac{1}{2}{V^\prime_C}^T B^{-1}V^\prime_C\,.
\ee
Thus if one of them is a minimum it is the deeper extremum, and
the other a saddle point. Hence the minimum is stable.
\item For a pair of extrema $C$ and $E$,
\be
V_E \,-\,V_C\,=\,\frac{1}{4}\,\left(w_E^2\,m^2_{C2}\,-\,v_C^2\,m^2_{E1}\,-\,z_C^2\,m^2_{E3}\right).
\ee
Both extrema can be simultaneously minima, and none of them is necessarily deeper
than the other. Stability of either extrema is thus not ensured.
\item For a pair of extrema $C$ and $F$,
\be
V_F \,-\,V_C\,=\,-\,\frac{1}{4}\,v_C^2\,m^2_{F1}\,\,=\,
\frac{v_C^2}{8\,\lambda_3\,w_C^2}\,m^2_{C1}\,m^2_{C3}.
\ee
Thus if one of them is a minimum it is the deeper extremum, and
the other a saddle point. Hence the minimum is stable.
\item For a pair of extrema $C$ and $G$,
\be
V_G \,-\,V_C\,=\,\frac{1}{4}\,\left(w_G^2\,m^2_{C2}\,-\,v_C^2\,m^2_{G1}\right).
\ee
Both extrema can be simultaneously minima, and none of them is necessarily deeper
than the other. Stability of either extrema is thus not ensured.
\item For a pair of extrema $D$ and $E$,
\be
V_E \,-\,V_D\,=\,-\,\frac{1}{4}\,\left(v_D^2\,m^2_{E1}\,+\,z_D^2\,m^2_{E3}\right)\,=\,
\frac{1}{2}{V^\prime_E}^T B^{-1}V^\prime_E\,.
\ee
Thus if one of them is a minimum it is the deeper extremum, and
the other a saddle point. Hence the minimum is stable.
\item For a pair of extrema $D$ and $F$,
\be
V_F \,-\,V_D\,=\,-\,\frac{1}{4}\,\left(v_D^2\,m^2_{F1}\,+\,w_D^2\,m^2_{F2}\right)\,=\,
\frac{1}{2}{V^\prime_F}^T B^{-1}V^\prime_F\,.
\ee
Thus if one of them is a minimum it is the deeper extremum, and
the other a saddle point. Hence the minimum is stable.
\item For a pair of extrema $D$ and $G$,
\be
V_G \,-\,V_D\,=\,-\,\frac{1}{4}\,v_D^2\,m^2_{G1}\,=\,
\frac{1}{2}{V^\prime_G}^T B^{-1}V^\prime_G\,.
\ee
Thus if one of them is a minimum it is the deeper extremum, and
the other a saddle point. Hence the minimum is stable.
\item For a pair of extrema $E$ and $F$,
\be
V_F \,-\,V_E\,=\,\frac{1}{4}\,\left(z_F^2\,m^2_{E3}\,-\,w_E^2\,m^2_{F2}\right).
\ee
Both extrema can be simultaneously minima, and none of them is necessarily deeper
than the other. Stability of either extrema is thus not ensured.
\item For a pair of extrema $E$ and $G$,
\be
V_G \,-\,V_E\,=\,\frac{1}{4}\,z_G^2\,m^2_{E3}\,\,=\,
-\,\frac{z_G^2}{8\,\lambda_2\,w_G^2}\,m^2_{G2}\,m^2_{G3}.
\ee
Thus if one of them is a minimum it is the deeper extremum, and
the other a saddle point. Hence the minimum is stable.
\item For a pair of extrema $F$ and $G$,
\be
V_G \,-\,V_F\,=\,\frac{1}{4}\,w_G^2\,m^2_{F2}\,\,=\,
-\,\frac{w_G^2}{8\,\lambda_3\,z_G^2}\,m^2_{G2}\,m^2_{G3}.
\ee
Thus if one of them is a minimum it is the deeper extremum, and
the other a saddle point. Hence the minimum is stable.
\end{itemize}

\section{Potential with a global $U(1)$ symmetry}
\label{app:U1}

A very commonly used version of the complex singlet-doublet model includes
a global $U(1)$ symmetry \cite{Barger:2008jx}, instead of the discrete ones
we have been using. That model's potential has far less parameters, namely
only five of them. In terms of the doublet $\Phi$ and the complex singlet
$\chi$, the potential can be read off eq. \eqref{eq:Vcomp} and is given by
\ba
V_{U(1)} &=& \mu^2_{\Phi}|\Phi|^2\,+\,\mu^2_{\chi} |\chi|^2
\,+\, \frac{1}{2}\lambda_\Phi |\Phi|^4 \,+\, \frac{1}{2}\lambda_{\chi} |\chi|^4  \,+\, \lambda_{\Phi \chi}|\Phi|^2|\chi|^2
\label{eq:VU1}
\ea
with obvious renamings of some of the couplings for convenience. This model can now have three
types of extrema:
\begin{itemize}
\item Type $A$, where $\langle\Phi\rangle = v_A/\sqrt{2}$,  $\langle\chi\rangle = 0$. The global
$U(1)$ symmetry is unbroken. Apart from the Goldstone bosons, there are two non-zero
masses (the singlet's components are mass degenerate), given by
\be
m^2_{A1} = \lambda_\Phi v_A^2\;\;\; , \;\;\; m^2_{A2} = \mu^2_{\chi} +
\frac{1}{2}\lambda_{\Phi \chi} v_A^2\,.
\ee
\item Type $B$, where $\langle\Phi\rangle = v_B/\sqrt{2}$,  $\langle\chi\rangle = w_B/\sqrt{2}$.
The global $U(1)$ symmetry is broken and a zero-mass scalar is generated (on top of the three
Goldstone bosons). One of the singlet's components mixes with $\varphi_3$ and there
are two scalars with masses
\be
m^2_{B1,2}\,=\,\frac{1}{2}\left[ \lambda_\Phi\,v_B^2\,+\,\lambda_\chi\,w_B^2\,\pm
\sqrt{\left(\lambda_\Phi\,v_B^2\,-\,\lambda_\chi\,w_B^2\right)^2\,+\,4\,
\lambda_{\Phi \chi}^2\,v_B^2\,w_B^2}\,\right]\,.
\ee
\item Type $C$, where only the singlet acquires a vev: $\langle\Phi\rangle = 0$,
$\langle\chi\rangle = w_C/\sqrt{2}$. Electroweak symmetry is unbroken, but the
singlet's vev breaks the global $U(1)$ symmetry, generating a massless scalar.
All of the doublets components are degenerate in mass (with mass given by
$m_{C1}$ below), and there is a second non-zero mass. The mass expressions are
\be
m^2_{C1} = \mu^1_{\chi} + \frac{1}{2}\lambda_{\Phi \chi} w_C^2\;\;\; , \;\;\;
m^2_{C2} = \lambda_\chi w_C^2\,.
\ee
\end{itemize}

Due to the form of the potential \eqref{eq:VU1}, we can always, without loss
of generality, assume that all of the vevs above are real. Applying the same
methods which led to the results of section \ref{sec:min}, it is simple to arrive
at the following expressions for the relative depths of each extrema:
\begin{itemize}
\item For simultaneous extrema $A$ and $B$,
\be
V_B - V_A \,=\, \frac{1}{4} w_B^2 m^2_{A2}\,=\, -\frac{w_B^2}{8 \lambda_\Phi v_B^2} m^2_{B1} m^2_{B2}\,.
\ee
Thus if $A$ ($B$) is a minimum, then it is deeper than $B$ ($A$), which is necessarily
a saddle point.
\item For simultaneous extrema $A$ and $C$,
\be
V_C - V_A \,=\, \frac{1}{2}\,\left(\frac{\mu_\Phi^4}{\lambda_\Phi}\,-\,\frac{\mu_\chi^4}{\lambda_\chi}\right)\,.
\ee
$A$ and $C$ can be simultaneously minima, and stability of a minimum of type
$A$ is not guaranteed. The expression above shows that when an extremum of type $C$
exists, to ensure $A$ is the global minimum it is necessary to require that
$\lambda_\chi\mu_\Phi^4\,>\,\lambda_\Phi\mu_\chi^4$.
\item For simultaneous extrema $B$ and $C$,
\be
V_C - V_B \,=\, -\frac{1}{4} v_B^2 m^2_{C1}\,=\, -\frac{v_B^2}{8 \lambda_\chi w_B^2} m^2_{B1} m^2_{B2}\,.
\ee
Thus if $B$ ($C$) is a minimum, then it is deeper than $C$ ($B$), which is necessarily
a saddle point.
\end{itemize}

The vacuum structure is therefore much simpler than the model with discrete $Z_2$ symmetries,
although the stability of a vacuum of type $A$ (the only acceptable one, seeing as how
electroweak symmetry is broken and there are no physical massless scalars) is not
guaranteed.

\bibliography{references}

\end{document}